\documentclass[preprintnumbers,amsmath,amssymb,floatfix,11pt,prd,onecolumn,superscriptaddress,nofootinbib]{revtex4}
\usepackage[utf8]{inputenc}
\usepackage{diagbox}
\usepackage{latexsym}
\usepackage{epsfig}
\usepackage{epstopdf,float}
\usepackage{graphicx}
\usepackage{amssymb}
\usepackage{amsmath}
\usepackage{dcolumn}
\usepackage{bm}
\usepackage{color}
\usepackage{comment}
\usepackage[shortlabels]{enumitem}
\usepackage{subfigure}
\usepackage{multirow}
\usepackage{xcolor}
\begin{document}
\title{\bf Shadows and Polarization Images of a Four-dimensional Gauss-Bonnet Black Hole Irradiated by a Thick Accretion Disk}
\author{Xiao-Xiong Zeng}
\altaffiliation{xxzengphysics@163.com} \affiliation{College of
Physics and Electronic Engineering, Chongqing Normal University,
Chongqing $401331$, People's Republic of China}
\author{Huan Ye}
\affiliation{School of
Material Science and Engineering, Chongqing Jiaotong University,
\\Chongqing 40007, People's Republic of China}
\author{Muhammad Israr Aslam}
\altaffiliation{mrisraraslam@gmail.com,
israr.aslam@umt.edu.pk}\affiliation{Department of Mathematics,
School of Science, University of Management and Technology,
Lahore-$54770$, Pakistan.}
\author{Rabia Saleem}
\altaffiliation{rabiasaleem@cuilahore.edu.pk}\affiliation{Department
of Mathematics, COMSATS University Islamabad, Lahore Campus,
Lahore-$54000$ Pakistan.}

\begin{abstract}
We adopt a general relativistic ray-tracing approach to study the
shadows and polarization images of spherically symmetric
Gauss-Bonnet (GB) black holes enveloped by geometrically thick
accretion flows. Specifically, we adopt a phenomenological RIAF-like
model and an analytical Hou disk model. In the RIAF-like model,
increasing the GB coupling parameter $\lambda$ reduces both the size
and brightness of the higher-order image, while increasing $\theta$
alters the shape of the higher-order image and obscures the
horizon's outline. The main difference between isotropic and
anisotropic emission is that the latter produce distortion of the
high-order image in the vertical direction, leading to an elliptical
morphology. For the Hou disk model, due to specific regions being
geometrically thinner with the conical approximation, the high-order
images are narrower with the increase in $\lambda$ than the RIAF
model. While increasing $\theta$ enhances the brightness of the
direct images outside the higher-order images, but hardly changes the
size of the higher-order images, which is in sharp contrast to the
RIAF model. Meanwhile, the Hou disk produces polarization patterns
that trace the brightness configuration and are affected by
$\lambda$ and $\theta$, reflecting the intrinsic structure of
spacetime. These results illustrate that intensity and polarization
in thick-disk models provide probes of GB black holes and
near-horizon accretion dynamics.
\end{abstract}
\date{\today}
\maketitle

\section{Introduction}
The evidence of black holes existence at galactic centers has been
strongly supports by accumulating astronomical observations. Their
existence is now confirmed with the help of multiple independent
observations, detection of gravitational waves from a binary merger
of black holes, as observed by the LIGO/Virgo collaboration
\cite{GB1}, and the horizon-scale images of black hole shadows
released by the Event Horizon Telescope (EHT) \cite{GB2,GB3}. The
recent advancements in EHT polarization observations \cite{GB4,GB5},
revealing the intricate physics about the magnetized plasma and
radiation in the vicinity of black holes, offers unprecedented
insights into jet formation mechanisms. These developments has
sparked widespread interest in the modeling of black hole accretion
disks in extreme astrophysical scenario, particularly close to the
event horizon, within both General Relativity (GR)
\cite{Hou:2023bep,Zhang:2024lsf} and extended gravitational theories
\cite{GB6,GB7,GB8}.

It is widely acknowledged that, GR is well-tested in the weak
gravitational field \cite{GB9}, whereas the strong gravitational
field is far less constrained, departure considerable for modified
theories of gravity \cite{GB10,GB11}. Among the plethora of modified
theories of gravity, the Einstein Gauss-Bonnet (EGB) theory is a
promising candidate to investigate the existence of black hole
solutions distinct from those of GR \cite{GB12}. In four-dimensional
spacetime, the Lagrangian of EGB term is the topologically
invariant. And thus in order to consider the dynamical impact of EGB
gravity, one is generically need to work in higher dimensions
\cite{GB13,GB14}. Glavan and Lin are introduced EGB extended theory
in four-dimensional spacetime by rescaling the EGB coupling
parameter $\lambda$, such that
$\lambda\rightarrow\frac{\lambda}{D-4}$ and subsequently taking the
limit $D\rightarrow 4$ \cite{GB12}. However, there are some
criticism on EGB theory, as it is mentioned out that this is not
ideal with the initial regularization scheme \cite{GB15,GB16,GB17}.
In this regard, Aoki et al., \cite{GB18} have introduced a
consistent four-dimensional EGB theory using the ADM decomposition
of the spacetime. Consequently, the proposed theory serve as a
consistent realization of the four-dimensional EGB theory where
there exist two dynamical degrees of freedom by breaking the
temporal diffeomorphism invariance. Hence, the cosmological and
black hole solutions naively defined in \cite{GB12} can be accounted
as exact solutions in \cite{GB18}. The background black hole
solution considered here also satisfies the well-defined theoretical
framework of \cite{GB18}. As a consequence, during the recent years
various significant properties of black hole solution in
four-dimensional GB has been investigated, for instance see Refs
\cite{GB19,GB20,GB21,GB22,GB23,GB24,GB25}. Specifically, some
authors have been investigated the black hole thermodynamics
\cite{GB26}, quasinormal modes \cite{GB27}, shadows and light rings
\cite{Zeng:2020dco,GB28} and Einstein rings with wave optics
\cite{GB29} etc. These investigations draw particular attention to
EGB gravity, specifically, this framework is closely connected to
the effective theories of superstrings, which are incorporated
through the GB coupling constant. Additionally, the shadows of black
holes also investigated in other modified theories of gravity
through a variety of accretion models, such as spherical accretion
models \cite{th25}, thin accretion disk models \cite{gkref26,sd38},
shadows of rotating black hole with celestial light sphere and thin
accretion disk \cite{sd34,sd35,sdn1} and wave optics
\cite{sd22,sd23,israr1} as well.

It is well-known that, supermassive black holes accrete hot,
magnetized plasma, giving rise to luminous disks in which thermal
synchrotron emission likely dominates the observed image frequencies
\cite{GB30}. The observational characteristics of accreting disk
encode both gravitational and intrinsic plasma effects, including
the electron distribution, temperature, and magnetic field geometry.
As suggested in \cite{GB31,GB32}, based on the dominant effects of
gravity, accretion flows near supermassive black holes may become
geometrically thick and optically thin, owing to the suppression of
vertical cooling and compression of matter \cite{GB33}. In this
regard, it becomes necessary to consider the comprehensive
configurations of streamlines, particle number density, electron
temperature, and magnetic field structure \cite{GB34}. Considering
the framework of geometrically thick accretion disks, various
studies impose the radiatively inefficient accretion flow (RIAF)
model with relatively low radiation efficiency, in which the
vertically averaged electron number density and temperature
typically follow a power-law distribution with radius
\cite{GB35,GB36,GB37,GB38}. This phenomenological model are
well-fitted with general relativistic magnetohydrodynamic (GRMHD)
simulations, which has been useful in generating the sub-millimeter
spectra and overall shape of image like as M$87^{\ast}$ \cite{GB39}.
In \cite{GB40} authors utilized a model, which is provide the
magnetized torus solution. Actually, this model is describe the
toroidal configuration, which was frequently considered as the
initial condition for GRMHD simulations. Notably, some
investigations have also considered the images of the torus
\cite{GB41,GB42}.

In literature, many traditional accretion disk models emphasize
nearly circular particle motion, and large-scale characteristics. In
contrast, recent horizon-scale observations make it possible to
probe accreting behavior in the vicinity of the black hole event
horizon. In this perspectives, Hou et al. \cite{GB43,GB44} proposed
a self-consistent analytical description of horizon-scale accretion
within the fabric of GRMHD, so-called ballistic approximation
accretion flow (BAAF) model. The central assumption of this scheme
is that gravitational forces dominate the fluid acceleration close
to the horizon. Under this approximation, the model provides
explicit expressions for key thermodynamic variables and magnetic
field structures, leading to a physically motivated description of
the morphology and dynamics of geometrically thick accretion flows.
As a result, the BAAF framework offers a powerful tool for studying
polarization signatures originating from the near-horizon region in
the vicinity of black holes. Polarization images are valuable tools
for discussing the dynamics of matter and magnetic field geometry in
the vicinity of black holes \cite{th30}. In this perspectives, this
simplified model successfully reproduced the observed configuration
of electric vector position angles and the relative polarization
intensity in the polarization image of M$87^{\ast}$. Within the
context of Schwarzschild black hole, Beloborodov derived a light-ray
approximation \cite{th31}. Moreover, the authors in \cite{th33}
developed a simplified model with an equatorial emission source,
produced the corresponding polarization images, and investigated the
geometric effects of black hole spin on photon parallel transport.
Subsequently, the polarization images of other black hole models and
horizonless ultra-compact objects have also been investigated in
\cite{young1,th34,th36,young2,th37,th38,th39,young3,zengnew1,zengnew2,GB45,GB46}.

Based on these studies, the present analysis aims to investigate the
optical characteristics of a static black hole within the context of
EGB gravity. Our main objective is to examine the influence of key
physical parameters on the black hole shadow and polarization
signatures. To this end, we consider two representative accretion
flow models, such as the phenomenological RIAF and the analytical
BAAF. The images are produced using a general relativistic radiative
transfer (GRRT) approach, and the resulting intensity and
polarization structure are analyzed by varying the GB coupling
parameter $\lambda$, the observer's inclination, and the observing
frequency. Additionally, we incorporate the effects of anisotropic
synchrotron emission. This mechanism provides a robust framework for
probing polarization properties in the near-horizon region of black
holes.

The organization of this paper is settled as follows: In sec. {\bf
II}, we briefly define the background of GB black hole including the
definition of null geodesics. In sec. {\bf III}, we define the
synchrotron radiation framework and the general relativistic GRRT
methodology. In the same section, we investigate the imaging
properties with the RIAF and the Hou disk models, with the effects
of emission anisotropy and distinct flow dynamics. We presents the
basic formalism of polarization imaging in sec. {\bf IV}, while sec.
{\bf V} is devoted to investigate the corresponding intensity and
polarization patterns. Finally, we end the paper with a conclusion
in last section. Throughout this analysis, we work under the
geometric unit system where $G = c = 1$.

\section{Gauss-Bonnet Black Hole and Null Geodesics}
The Einstein-Hilbert action with an additional GB term is defined as
follows:
\begin{equation}\label{action1}
\mathcal{I}=\frac{1}{16 \pi G}\int
d^4{x\sqrt{-g}}\big(R+\lambda(R_{\mu\nu\delta\eta}R^{\mu\nu\delta\eta}-4R_{\delta\eta}R^{\delta\eta}+R^{2})\big),
\end{equation}
in which $R$ is the Ricci scalar, by rescaling the GB coupling
constant $\lambda\rightarrow\frac{\lambda}{D-4}$ and taking the
limit $D\rightarrow 4$ in the GB term, one can obtain the
four-dimensional spherically symmetric black hole metric in GB
gravity can be expressed as \cite{Zeng:2020dco}
\begin{equation}
ds^{2} = -F(r) dt^{2} + \frac{dr^{2}}{F(r)} + r^{2} (d\theta^{2} +
\sin^{2}\theta d\phi^{2}),
\end{equation}
where
\begin{equation}
F(r) = 1 + \frac{r^{2}}{2\lambda} \left(1 - \sqrt{1 + \frac{8\lambda
M}{r^{3}}} \right).
\end{equation}
Here $M$ is the black hole mass. For the sake of convenience in
calculations, $M$ is set to $1$ throughout the following text. In
order to investigate the motion of massless particles in the
vicinity of GB black hole, one can define the Lagrangian formalism
as
\begin{eqnarray}\label{pn1}
\mathcal{L}=\frac{1}{2}g_{\alpha\beta}\dot{x}^{\alpha}\dot{x}^{\beta}=
\frac{1}{2}\big(-F(r)\dot{t}^{2}+\frac{\dot{r}^{2}}{F(r)}+r^{2}(\dot{\theta}^{2}+\sin^{2}\theta
\dot{\phi}^{2})\big),
\end{eqnarray}
here $\dot{x}^{\alpha}$ is the four-velocity of the photon. Since
the GB black hole is spherically symmetric, so we assume photons
moving in the equatorial plane $\theta=\pi/2$. Along a null
geodesic, a photon possesses two conserved quantities, such as
energy $E=-p_{t}$ and the angular momentum $L=p_{\phi}$. Combining
these with the null normalization constraint
$u^{\alpha}u_{\beta}=0$, one can define the radial equation of
motion:
\begin{eqnarray}\label{pn2}
\bigg(\frac{dr}{d\eta}\bigg)^{2}=-V_{eff}(r),
\end{eqnarray}
where
\begin{eqnarray}\label{pn3}
V_{eff}(r)=E^{2}\bigg(1-\frac{F(r)}{r^2}\hat{b}^{2}\bigg),
\end{eqnarray}
here $V_{eff}$ and $\hat{b}=\frac{L}{E}$, known as effective
potential and impact parameter, respectively. The circular orbits
can then be evaluated by solving the following equation as:
\begin{equation}
\frac{\partial V_{eff}(r)}{\partial r} = 0. \label{peq:effs}
\end{equation}

\section{Non-Polarized Imaging}
In the non-polarized case, this section considers two geometrically
thick, optically thin accretion disk models: namely, the
phenomenological model and the HOU disk model.

\subsection{Phenomenological model}
The phenomenological model is based on the inefficient accretion
flow models by Yuan et al. \cite{Yuan:2003dc} and further developed
by Broderick et al. \cite{Broderick:2010kx}. In this model, the
number density and temperature distribution of non-thermal electrons
can be described as:
\begin{equation}
n_e = n_h \left( \frac{r}{r_h} \right)^2 \exp \left(
-\frac{z^2}{2(\alpha R)^2} \right),
\end{equation}
and
\begin{equation}
T_e = T_h \left( \frac{r}{r_h} \right),
\end{equation}
respectively, where $R = r \sin \theta$ denotes the radius in
cylindrical coordinates, and $z = r \cos \theta$ represents the
height from the equatorial plane $\theta = \pi/2$ in cylindrical
coordinates. Here, $n_h$, $\alpha$, and $T_h$ are constants, and
$r_h$ is the outer horizon. The magnetic field strength can be
defined through the cold magnetization parameter:
\begin{equation}
\sigma = \frac{B^2}{\rho} = \frac{B^2}{n_e (m_p c^2)},
\end{equation}
which gives:
\begin{equation}
B = \sqrt{\sigma \rho},
\end{equation}
where the dimensionless quantity $\rho = n_e (m_p c^2)$ represents
the fluid mass density. The parameter $\sigma$ is a constant, and
for the accretion disk model considered here, its magnitude is on
the order of $\sigma \sim 0.1$. For isotropic radiation, only the
magnetic field strength is considered, disregarding the direction of
the magnetic field. Thus, the angle between the magnetic field and
the emitted photons is not taken into account. For anisotropic
radiation, the direction of the magnetic field can be considered as
a mixture of toroidal and poloidal components, specifically
expressed as:
\begin{equation}
B^\mu \sim (l, 0, A, 1), \label{eq7}
\end{equation}
where $A$ is an adjustable parameter, set to $0$ in this paper, and:
\begin{equation}
l = -\frac{u_\phi}{u_t}, \quad u_\nu = g_{\mu\nu} u^\mu = (u_t, u_r,
u_\theta, u_\phi).
\end{equation}
This form of the magnetic field satisfies the orthogonality
condition with the fluid four-velocity: $u^\mu B_\mu = 0$. After
determining the basic properties of the accretion flow, we consider
its motion. Typically, we can consider free-fall motion, circular
motion around the black hole, or even a combination of both. In this
paper, we mainly consider the fluid in free-fall. Assuming the fluid
is at rest at infinity, the corresponding four-velocity can be
expressed as:
\begin{eqnarray}
u_t = -g^{tt}, \quad u^r = -\sqrt{-(1 + g^{tt})g^{rr}}.
\end{eqnarray}
The four-velocity must be a timelike vector throughout the entire
spacetime, requiring $g^{tt} \leq -1$. After establishing the basic
properties and motion of the accreting matter, we will discuss the
motion of electrons via the radiative transfer equation. The
covariant form of the radiative transfer equation for non-polarized
light is \cite{Gold:2020iql}:
\begin{equation}
\frac{d}{d\omega}\mathcal{I} = \mathcal{J} - \alpha
\mathcal{I},\label{eqdi}
\end{equation}
where $\mathcal{I}$, $\mathcal{J}$, and $\alpha$ are generalized
invariants, related to the actual physical quantities by:
\begin{equation}
\mathcal{I} = \frac{I_\nu}{\nu^3}, \quad \mathcal{J} =
\frac{j_\nu}{\nu^2}, \quad \alpha =
\nu\alpha_\nu.\label{eqi_j_alpha}
\end{equation}
Here, $I_\nu$ is called the specific intensity, defined as:
\begin{equation}
I_{\nu} = \frac{dE}{dA\,dt\,d\Omega\,d\nu} =
\frac{h^{4}\nu^{3}}{c^{2}} \frac{dN}{d^{3}r\,d^{3}p}.
\end{equation}
Its physical meaning is the energy $dE$ transported by
electromagnetic waves through an area $dA$, within a solid angle
$d\Omega$, and a frequency interval $d\nu$, during time $dt$.
Considering $E = N h \nu$, where $N$ is the number of photons
\begin{equation}
f = \frac{dN}{d^3 r d^3 p},
\end{equation}
is the number density distribution function of photons in phase
space $(\mathbf{r}, \mathbf{p})$. Since both $N$ and $d^3 r d^3 p$
are generalized invariants, this distribution function is also a
generalized invariant. And $j_\nu$ is called the emissivity, which
is defined as:
\begin{equation}
j_\nu = \frac{dE}{d^3 r dt d\nu d\Omega} = \frac{h^4 \nu^3}{c^3}
\frac{dN}{dt d^3 r d^3 p},
\end{equation}
while $\alpha_\nu$ is called the absorption coefficient, has the
following expression
\begin{equation}
\alpha_\nu = n \sigma_\nu,
\end{equation}
where $\sigma_\nu$ is the photon absorption cross-section, and $n$
is the number density of the medium. To convert from geometric units
to CGS units, we can multiply a coefficient before the affine
parameter in equation (\ref{eqdi}):
\begin{equation}
\frac{d}{d\omega} \rightarrow \frac{1}{C} \frac{d}{d\omega}, \quad C
= \frac{r_g}{\nu_0},
\end{equation}
where $r_g = GM/c^2$ is the unit length, and $\nu_0$ is the
frequency of the real photon at infinity. Equation (\ref{eqdi}) then
becomes:
\begin{equation}
\frac{1}{C} \frac{d}{d\omega}\mathcal{I} = \mathcal{J} - \alpha
\mathcal{I},
\end{equation}
and its solution is:
\begin{equation}
I_\nu = g^3 I_{\nu_0} + r_g \int_{\omega_0}^{\omega} d\omega' g^2
j_\nu(\omega') \exp\left( -r_g \int_{\omega'}^{\omega} d\omega''
\alpha_\nu(\omega'')/g \right),\label{eq_inu}
\end{equation}
where $g = \nu_0 / \nu$ is the redshift factor. Let the fluid
four-velocity be $u^\mu$ and the local magnetic field be $B^\mu$
(satisfying $B_\mu u^\mu = 0$), then:
\begin{equation}
g = \frac{k_\mu (\partial_t)^\mu}{k_\mu u^\mu} = \frac{k_t}{k_\mu
u^\mu} = \frac{-1}{k_\mu u^\mu}.
\end{equation}
The radiation coefficients $j_\nu$ and $\alpha_\nu$ in equation
(\ref{eqi_j_alpha}) are related to the considered radiation process.
This paper considers synchrotron radiation from electrons under
extreme relativistic conditions (using CGS units). In a plasma
system, synchrotron radiation primarily originates from electrons.
Its emissivity is:
\begin{equation}
j_\nu = \frac{\sqrt{3}e^3 B \sin\theta_B}{4\pi m_e c^2}
\int_0^\infty d\tau N(\tau) F\left( \frac{\nu}{\nu_s} \right),
\end{equation}
where $\tau = 1 / \sqrt{1 - \beta^2}$ is the Lorentz factor of the
charged particle, $N(\tau)$ is the electron distribution function,
and $F(x)$ is related to the modified Bessel function of the second
kind, $K_n(x)$:
\begin{equation}
F(x) = x \int_x^{\infty} dy K_{5/3}(y).
\end{equation}
Typically, different electron distributions correspond to different
radiation formulas. This paper considers a thermal distribution,
with the distribution function given by:
\begin{equation}
N(\tau) = n_e \frac{\tau^2 \beta}{\theta_e K_2(1/\theta_e)}
\exp\left( -\frac{\tau}{\theta_e} \right),
\end{equation}
where $n_e$ is the electron number density, $\theta_e = k_B T_e /
m_e c^2$ represents the dimensionless electron temperature, $T_e$ is
the thermodynamic temperature of the electrons, $e$ denotes electric
charge, $c$ is the speed of light, and $k_B$ is the Boltzmann
constant. For the extreme relativistic case, where $\beta \approx 1$
and $\theta_e \gg 1$, the asymptotic formula is $K_2(1/\theta_e)
\approx 2\theta_e^2$. Let $y = \tau/\theta_e$, then:
\begin{equation}
j_\nu = \frac{\sqrt{3}n_e e^3 B \sin\theta_B}{8\pi m_e c^2}
\int_0^\infty dy ~y^2 \exp(-y) F\left( \frac{\nu}{\nu_s}
\right),\label{eq_jnu}
\end{equation}
$\nu_s$ is the characteristic frequency:
\begin{equation}
\nu_s = \frac{3 e B \sin \theta_B \tau^2}{4\pi m c},
\end{equation}
and $\theta_B$ is the angle between $e^\mu_{(B)}$ and $e^\mu_{(k)}$:
\begin{equation}
\theta_B = \arccos \left( e^\mu_{(B)} \cdot e^\mu_{(k)} \right) =
\arccos \left[ \frac{g}{B} (B_\mu k^\mu) \right],
\end{equation}
where
\begin{eqnarray}
e^\mu_{(k)} = -\left( \frac{k^\mu}{u^\nu k_\nu} + u^\mu \right),
\quad e^\mu_{(B)} = \frac{B^\mu}{B}.
\end{eqnarray}
Let $s = (\nu/\nu_s)y^{2}$, then the emissivity becomes:
\begin{equation}
j_\nu = \frac{n_e e^2 \nu}{2\sqrt{3}c \theta_e^2} I(s), \quad s =
\frac{\nu}{\nu_c}, \quad \nu_c = \frac{3e B \sin\theta_B
\theta_e^2}{4\pi m_e c},
\end{equation}
where the parameter $s$ is the ratio of the emitted photon frequency
$\nu$ to the system's characteristic frequency $\nu_c$. The
dimensionless function is defined as:
\begin{equation}
I(s) = \frac{1}{s} \int_0^\infty y^2 \exp(-y) F\left( \frac{s}{y^2}
\right).
\end{equation}
This equation has no specific analytical formula and requires the
use of fitting functions. For isotropic radiation \cite{Leung:2011},
the fitting function is:
\begin{equation}
I(s) = \frac{4.0505}{s^{1/6}} \left( 1 + \frac{0.4}{s^{1/4}} +
\frac{0.5316}{s^{1/2}} \right) \exp\left( -1.8899s^{1/3}
\right),\label{eqI1}
\end{equation}
and for anisotropic radiation \cite{Mahadevan:1996cc}, the fitting
function is:
\begin{equation}
I(s) = 2.5651 \left( 1 + 1.92s^{-1/3} + 0.9977s^{-2/3} \right)
\exp\left( -1.8899s^{1/3} \right).\label{eqI2}
\end{equation}
Under the condition of a thermal electron distribution, the
absorption process follows Kirchhoff's law, which states that the
absorption coefficient must satisfy:
\begin{equation}
\alpha_{\nu} = \frac{j_{\nu}}{B_{\nu}}, \quad B_{\nu} =
\frac{2h\nu^{3}}{c^{2}} \cdot \frac{1}{\exp(h\nu/k_{B}T_{e}) -
1},\label{eq_alphanu_bnu}
\end{equation}
where $B_{\nu}$ represents the Planck black body radiation function.
Based on equations (\ref{eq_jnu}) and (\ref{eq_alphanu_bnu}), in
principle we can numerically solve equation (\ref{eq_inu}). Next, we
will consider the isotropic and anisotropic cases separately and
present the corresponding numerical results. Closely followed by
\cite{GBn1}, we adopt a zero angular momentum observer (ZAMO)
framework, where two-dimensional polar coordinate system are
introduced to see the optical image of black hole shadow on the
observer's screen, which is defined as
\begin{eqnarray}\label{s10}
X=\chi\cos\varphi, \quad Y=\chi\sin\varphi,
\end{eqnarray}
where $\chi$ denotes the radial distance from the image center and
$\varphi$ is the azimuthal angle measured counterclockwise on the
image plane.

\subsection{Phenomenological Model under Isotropic Radiation}
\begin{figure}[H]
\centering
\subfigure[$\lambda=0.01,\theta=17^\circ$]{\includegraphics[scale=0.37]{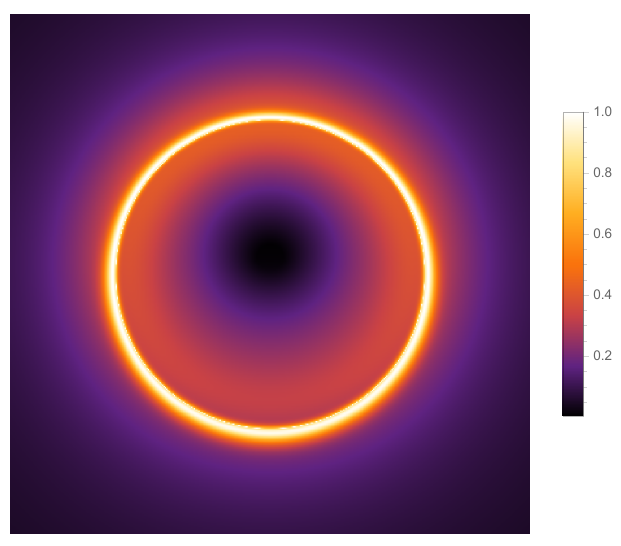}}
\subfigure[$\lambda=0.3,\theta=17^\circ$]{\includegraphics[scale=0.37]{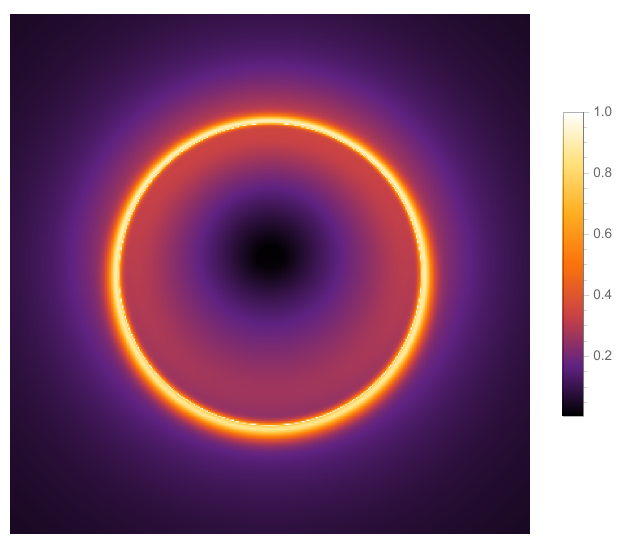}}
\subfigure[$\lambda=0.6,\theta=17^\circ$]{\includegraphics[scale=0.37]{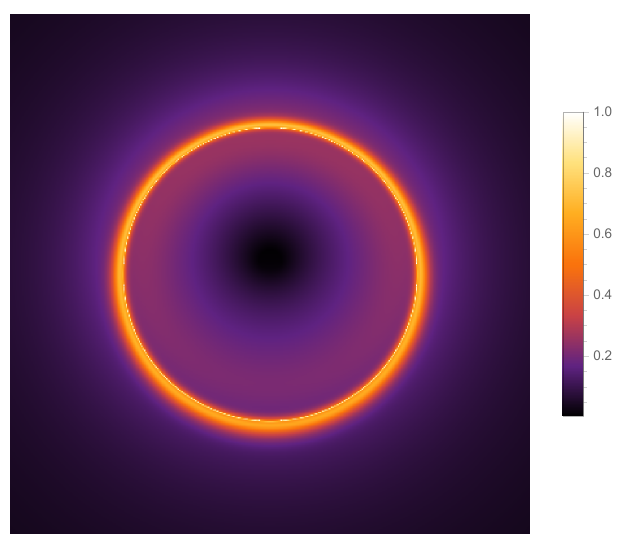}}
\subfigure[$\lambda=0.99,\theta=17^\circ$]{\includegraphics[scale=0.37]{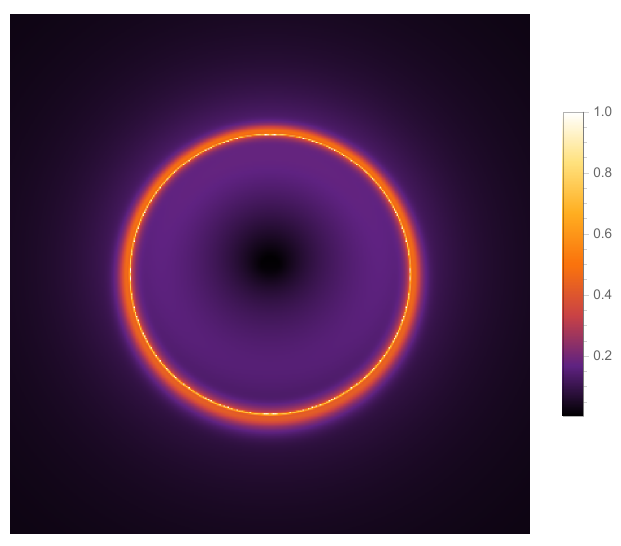}}
\subfigure[$\lambda=0.01,\theta=50^\circ$]{\includegraphics[scale=0.37]{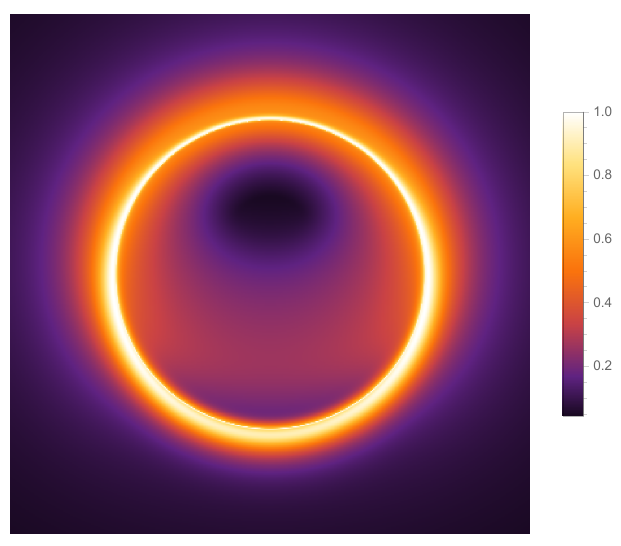}}
\subfigure[$\lambda=0.3,\theta=50^\circ$]{\includegraphics[scale=0.37]{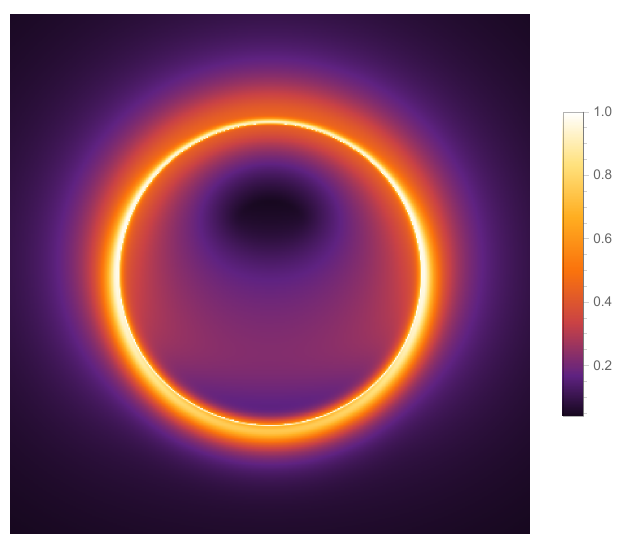}}
\subfigure[$\lambda=0.6,\theta=50^\circ$]{\includegraphics[scale=0.37]{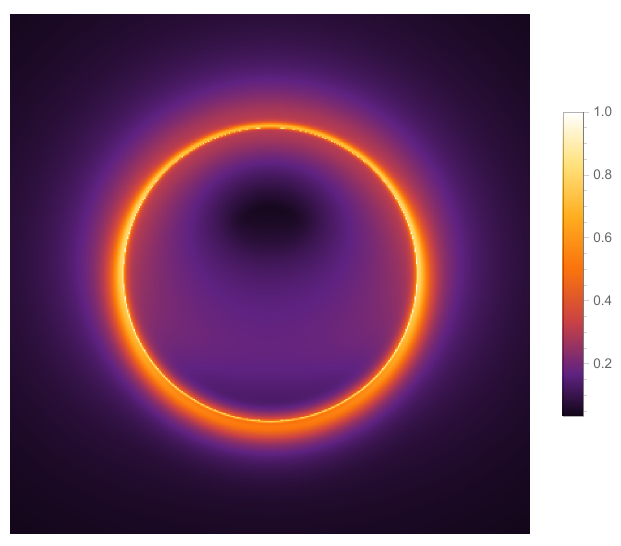}}
\subfigure[$\lambda=0.99,\theta=50^\circ$]{\includegraphics[scale=0.37]{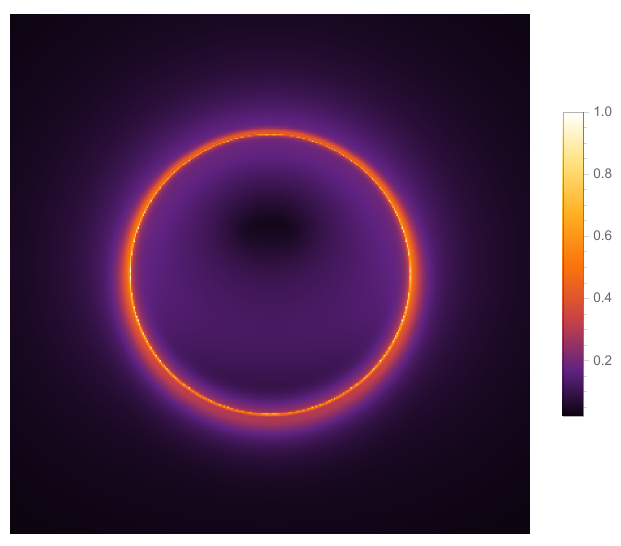}}
\subfigure[$\lambda=0.01,\theta=80^\circ$]{\includegraphics[scale=0.37]{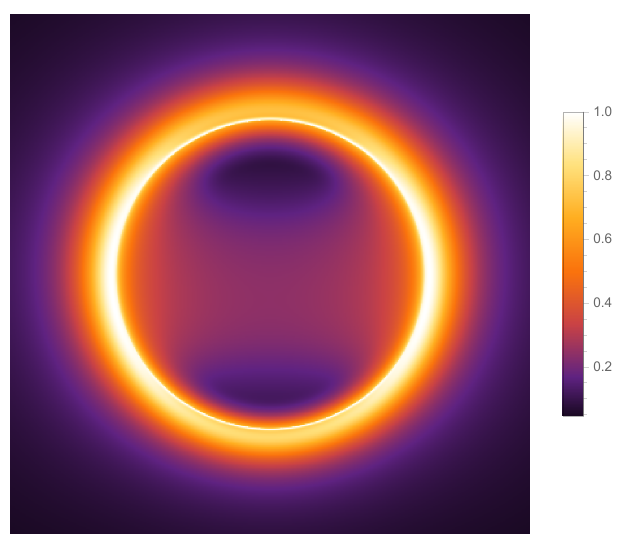}}
\subfigure[$\lambda=0.3,\theta=80^\circ$]{\includegraphics[scale=0.37]{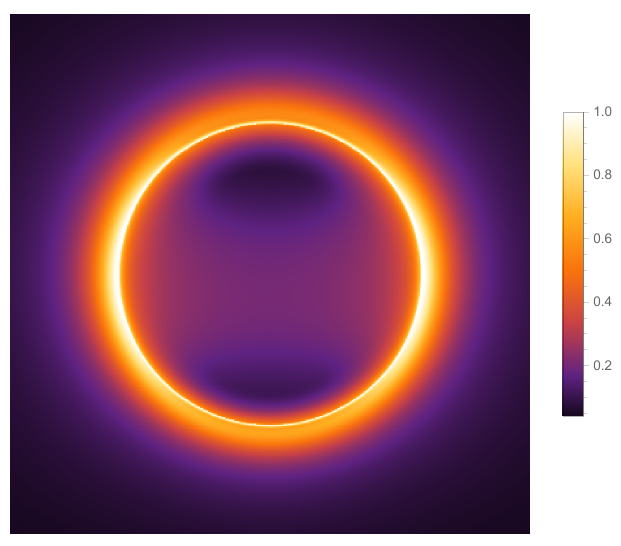}}
\subfigure[$\lambda=0.6,\theta=80^\circ$]{\includegraphics[scale=0.37]{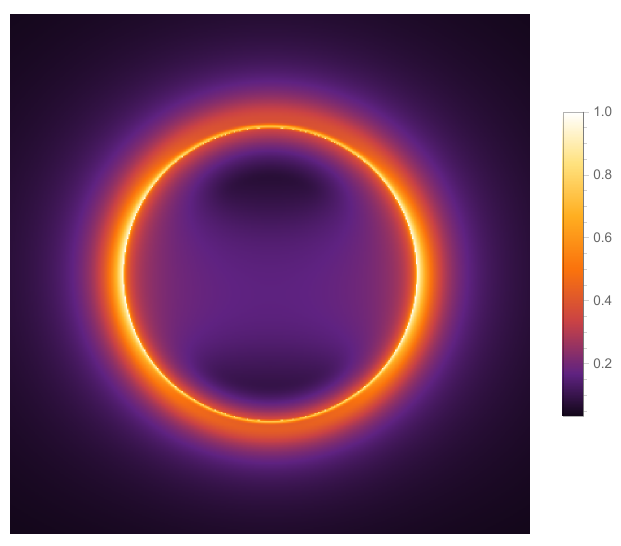}}
\subfigure[$\lambda=0.99,\theta=80^\circ$]{\includegraphics[scale=0.37]{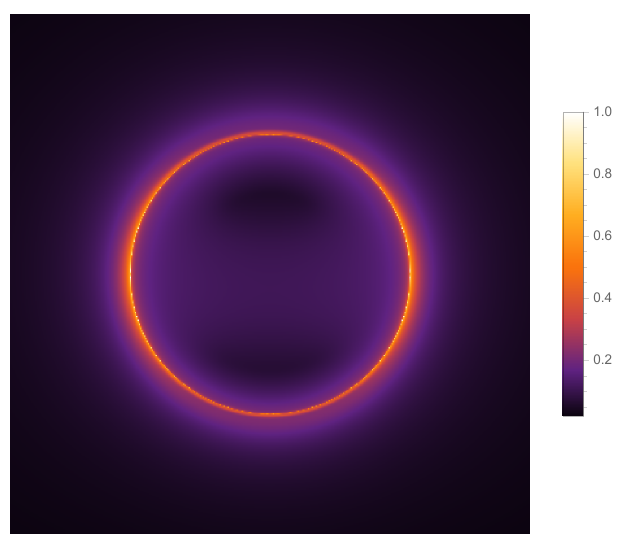}}
\caption{Shadow images of the black hole for the phenomenological
model under isotropic radiation.
   The accretion flow motion mode is infalling motion. From left to right,
   the parameter $\lambda$ takes values of $0.01,~0.3,~0.6,~0.99$ respectively.
   From top to bottom, the observation inclination angle $\theta$ takes
   values of $17^\circ,~50^\circ,~80^\circ$ respectively.
   The observer distance is fixed at $500M$, the field of view is $2^\circ$, and the observation frequency is $230$ GHz.}\label{figtong}
\end{figure}

Figure \textbf{\ref{figtong}} displays the shadow images of the
four-dimensional GB black hole under isotropic radiation in the
phenomenological model. We primarily investigate the case where the
accretion flow exhibits infalling motion. From left to right in the
images, the values of the parameter $\lambda$ are $0.01$, $0.3$,
$0.6$, and $0.99$, respectively. From top to bottom in the images,
the observer inclination angles $\theta$ are $17^\circ$, $50^\circ$,
and $80^\circ$, respectively. The observer distance is fixed at
$500M$, the field of view is $2^\circ$, and the observation
frequency is $230$ GHz. Observing Fig. \textbf{\ref{figtong}}, it
can be seen that a bright ring-like structure appears in all images,
corresponding to higher-order images, i.e., photons that orbit the
black hole once or multiple times before reaching the observer.
Outside this ring-like structure, the region where the intensity is
still non-zero corresponds to the primary image, i.e., photons that
travel directly from the accretion disk to the observer. Notably,
for all parameter values, a region with zero intensity appears
inside the higher-order images. This region originates from the
event horizon of the black hole. For geometrically thin accretion
disks, this region is referred to as the ``inner shadow'' and may be
captured by EHT \cite{Chael:2021rjo}. However, for the geometrically
thick accretion disk discussed in this paper, this region may be
obscured by radiation from outside the equatorial plane, making it
difficult to distinguish. Compared to thin disks, thick disks are
more physically realistic, indicating that direct imaging of the
black hole's event horizon remains challenging.

In Fig. \textbf{\ref{figtong}}, when $\theta = 17^\circ$ (first
row), the higher-order image exhibits an approximately ring-like
shape. As $\theta$ increases to $50^\circ$ (second row), the
higher-order image shifts downward on the screen. When $\theta$
further increases to $80^\circ$ (third row), the intensity on the
left and right sides of the higher-order image becomes significantly
stronger than that in the vertical directions. After fixing the
observation angle, we primarily study the influence of $\lambda$ on
the higher-order image. The results show that increasing $\lambda$
reduces the size of the higher-order image. In particular, as
$\lambda$ increases, the higher-order image becomes increasingly
dim. Interestingly, for $\theta = 80^\circ$ (third row), two dark
regions appear inside the higher-order image, with the upper region
slightly darker than the lower one. This phenomenon arises from
gravitational lensing effects. In summary, increasing $\lambda$
reduces both the size and brightness of the higher-order image,
while increasing $\theta$ alters the shape of the higher-order image
and obscures the horizon's outline.

\begin{figure}[H]
\centering
\subfigure[$85$~GHz]{\includegraphics[scale=0.5]{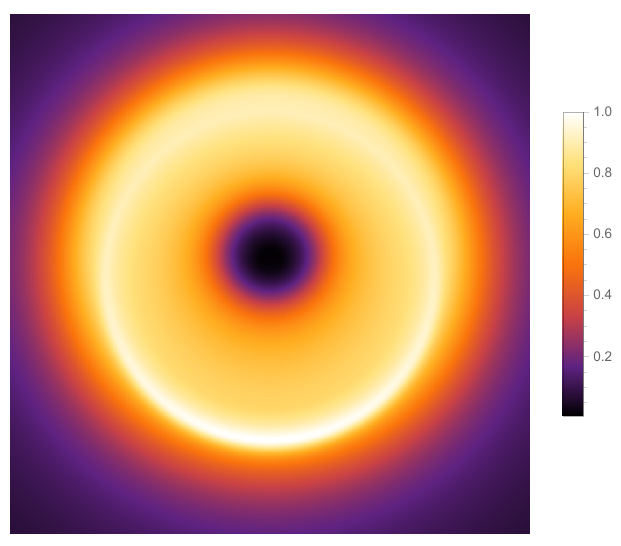}}
\subfigure[$230$~GHz]{\includegraphics[scale=0.5]{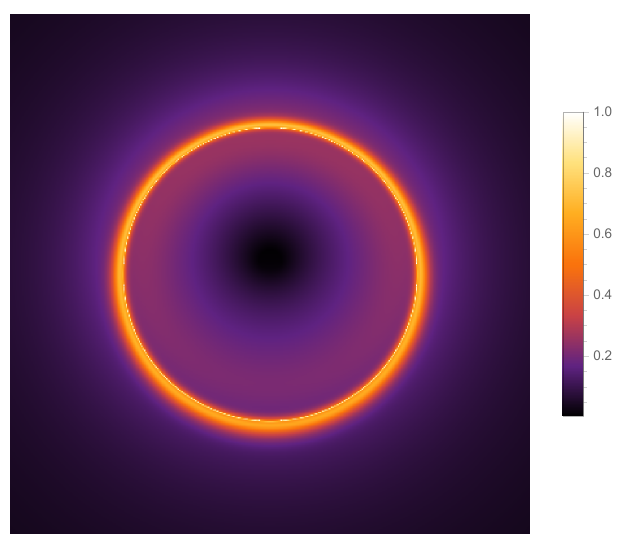}}
\subfigure[$345$~GHz]{\includegraphics[scale=0.5]{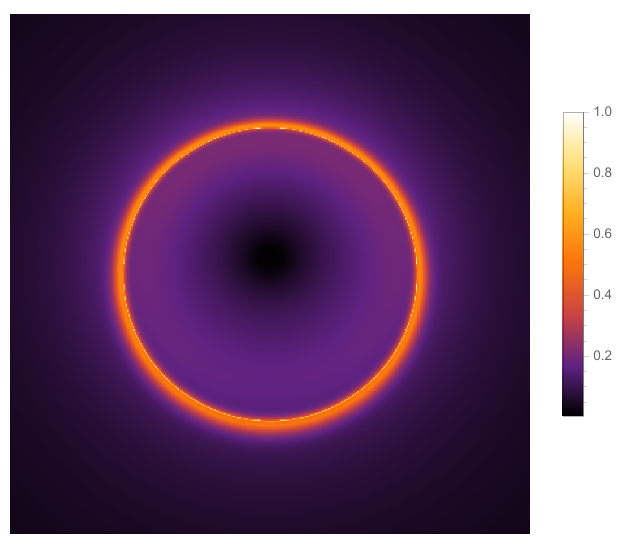}}
\subfigure[Horizontal
direction]{\includegraphics[scale=0.6]{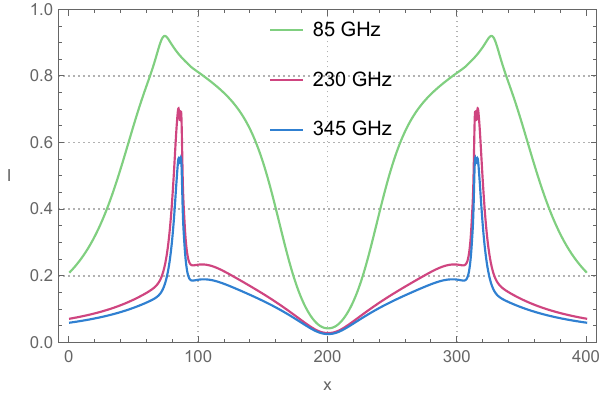}}
\subfigure[Vertical
direction]{\includegraphics[scale=0.6]{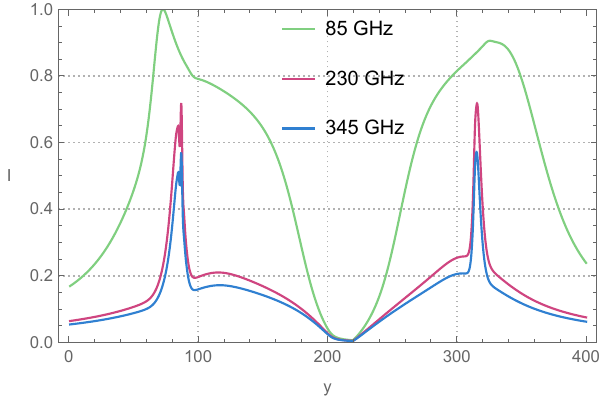}}
\caption{The first row of images shows: the black hole shadow images
under the phenomenological model with isotropic radiation. The
accretion flow motion mode is infalling motion. From left to right,
the observation frequencies are 85~GHz, 230~GHz, and 345~GHz. The
second row of images shows: intensity distribution profile plots at
different observation frequencies. The left plot is the horizontal
intensity distribution profile, and the right plot is the vertical
intensity distribution profile. The observer distance is fixed at
$500M$, the field of view is $2^\circ$, the observation inclination
is $17^\circ$, and $\lambda=0.6$.} \label{figtong_curve}
\end{figure}

Fig. \textbf{\ref{figtong_curve}} presents the characteristics of
the black hole shadow images under isotropic radiation for the
phenomenological model at different observation frequencies. The
first row of Fig. \textbf{\ref{figtong_curve}}, from left to right,
corresponds to observation frequencies of $85$ GHz, $230$ GHz, and
$345$ GHz, respectively. It can be observed that at $85$ GHz, the
horizon outline of the black hole is almost entirely obscured, and
the direct and higher-order images of the accretion disk become
difficult to distinguish. As the observation frequency increases,
the intensity decreases, making the higher-order image and the
horizon outline clearly visible.

In all intensity profile plots, two significant peaks can be
observed in each curve. The region outside the peaks corresponds to
the direct image of the accretion disk, formed by photons traveling
directly from the disk to the observer. The peaks themselves
represent the higher-order images, where photons have orbited the
black hole one or more times before reaching the observer. In the
first row of images, the higher-order image appears as a distinct
ring structure, which is a direct manifestation of strong
gravitational lensing. By comparing the three images and their
corresponding intensity profiles, it is evident that the intensity
of the ring structure decreases as the observation frequency
increases. However, the position of the higher-order image remains
unchanged across the three frequencies, because the gravitational
lensing effect of the black hole is independent of the frequency of
light.

\subsection{Phenomenological Model under Anisotropic Radiation}
\begin{figure}[H]
\centering
\subfigure[$\lambda=0.01,\theta=17^\circ$]{\includegraphics[scale=0.37]{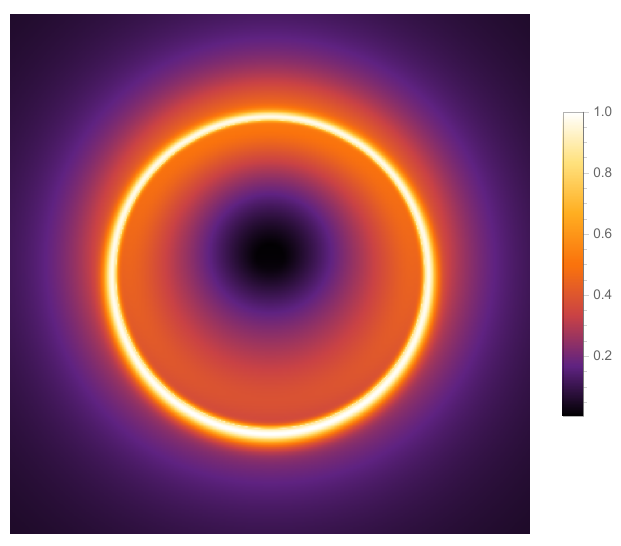}}
\subfigure[$\lambda=0.3,\theta=17^\circ$]{\includegraphics[scale=0.37]{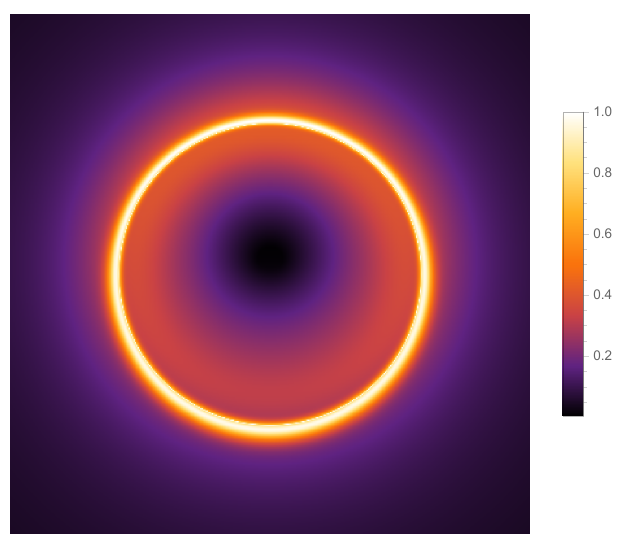}}
\subfigure[$\lambda=0.6,\theta=17^\circ$]{\includegraphics[scale=0.37]{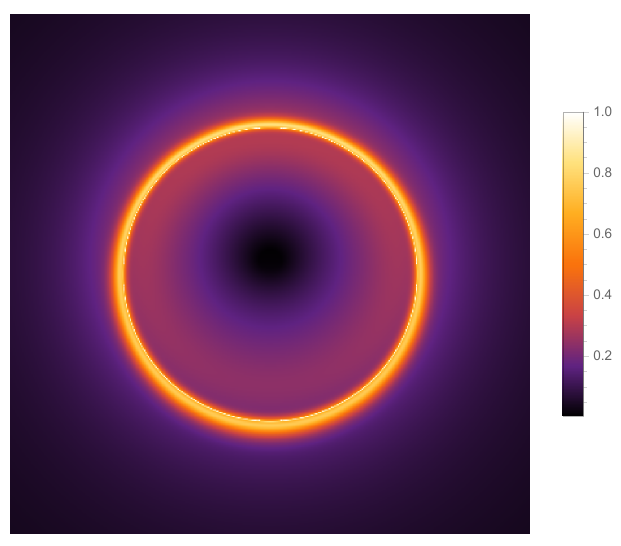}}
\subfigure[$\lambda=0.99,\theta=17^\circ$]{\includegraphics[scale=0.37]{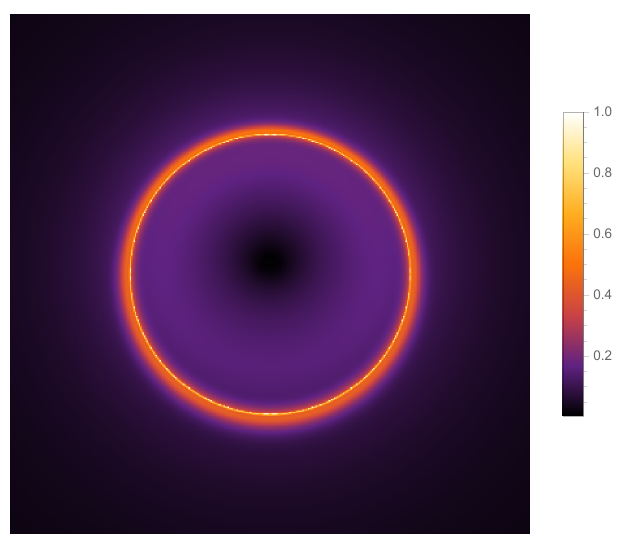}}
\subfigure[$\lambda=0.01,\theta=50^\circ$]{\includegraphics[scale=0.37]{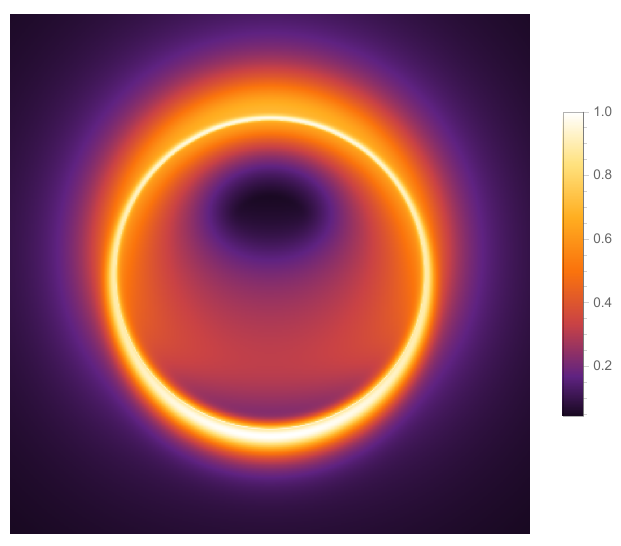}}
\subfigure[$\lambda=0.3,\theta=50^\circ$]{\includegraphics[scale=0.37]{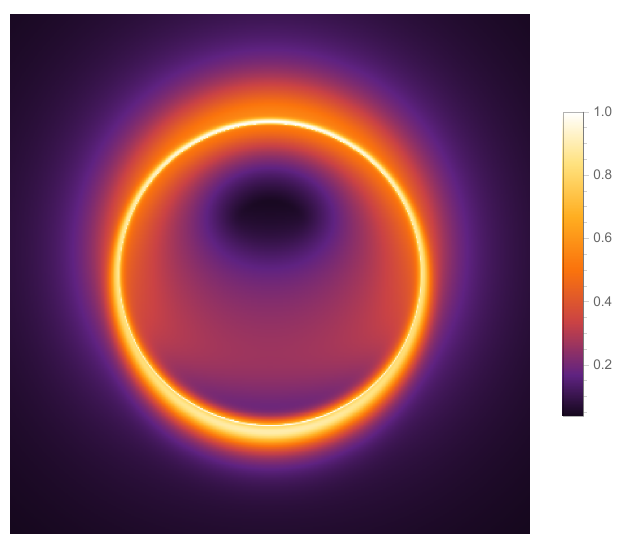}}
\subfigure[$\lambda=0.6,\theta=50^\circ$]{\includegraphics[scale=0.37]{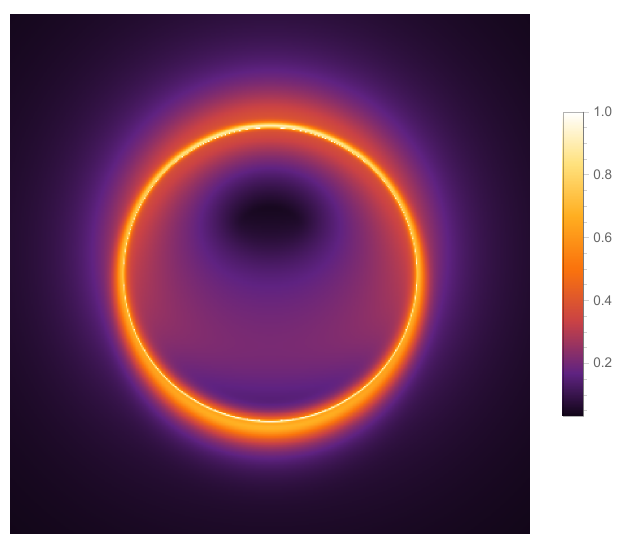}}
\subfigure[$\lambda=0.99,\theta=50^\circ$]{\includegraphics[scale=0.37]{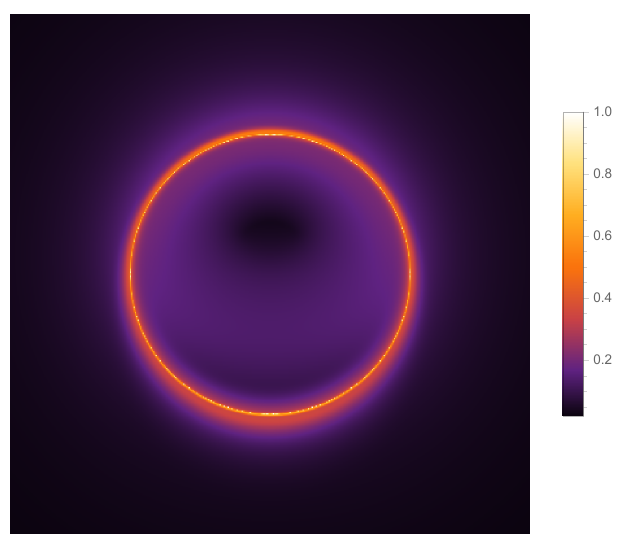}}
\subfigure[$\lambda=0.01,\theta=80^\circ$]{\includegraphics[scale=0.37]{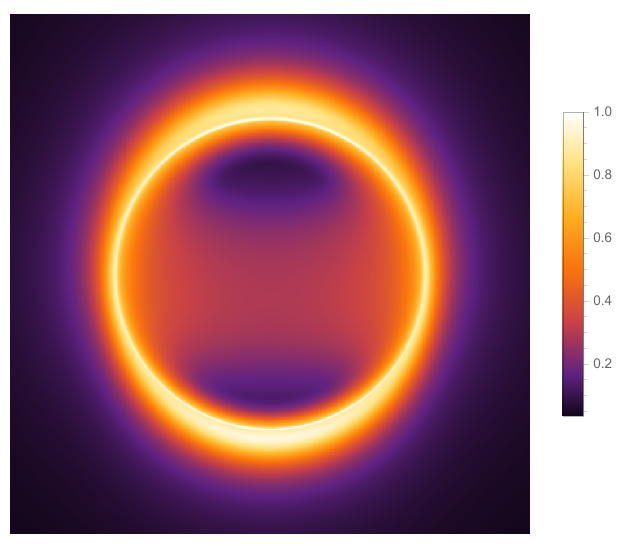}}
\subfigure[$\lambda=0.3,\theta=80^\circ$]{\includegraphics[scale=0.37]{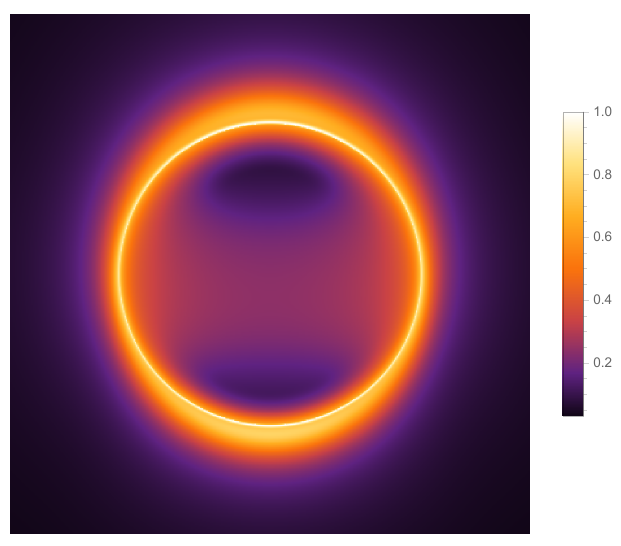}}
\subfigure[$\lambda=0.6,\theta=80^\circ$]{\includegraphics[scale=0.37]{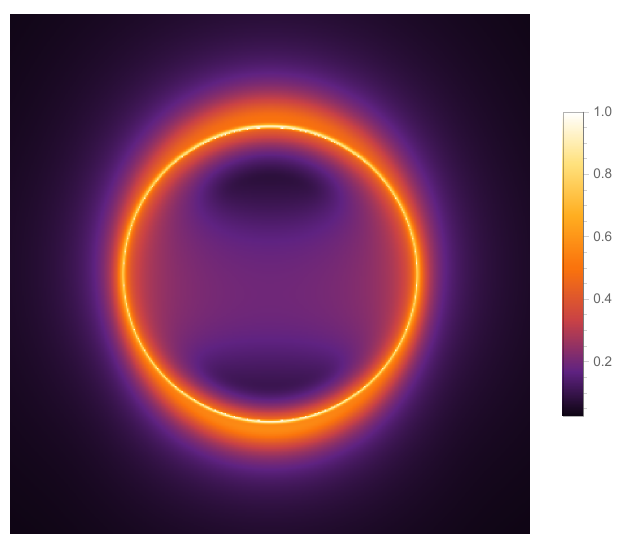}}
\subfigure[$\lambda=0.99,\theta=80^\circ$]{\includegraphics[scale=0.37]{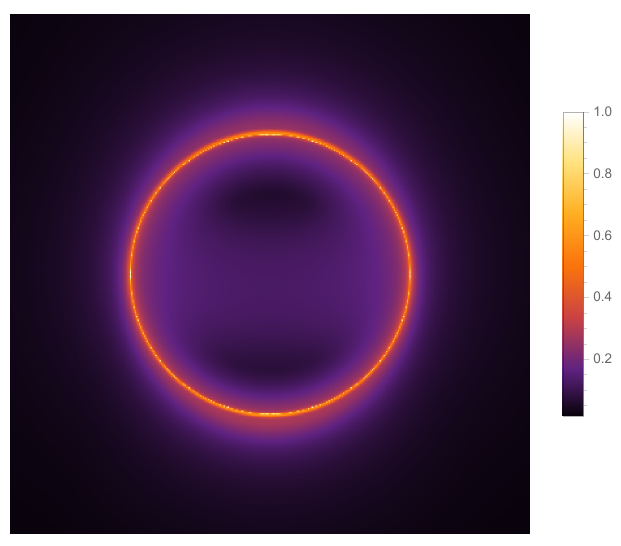}}
\caption{Black hole shadow images for the phenomenological model
under anisotropic radiation. The accretion flow motion mode is
infalling motion. From left to right, the parameter $\lambda$ takes
values $0.01,~0.3,~0.6,~0.99$, respectively. From top to bottom, the
observation inclination angle $\theta$ takes values
$17^\circ,~50^\circ,~80^\circ$, respectively. The observer distance
is fixed at $500M$, the field of view is $2^\circ$, and the
observation frequency is $230$ GHz.}\label{figyi}
\end{figure}

Figure \textbf{\ref{figyi}} presents the black hole shadow images of
the phenomenological model under anisotropic radiation, with an
observation frequency of 230 GHz and an accretion flow motion mode
of infalling motion. Observing Fig. \textbf{\ref{figyi}}, it can be
seen that the influence of $\lambda$ and $\theta$ on the black hole
shadow under anisotropic radiation is similar to that under
isotropic radiation. An increase in $\lambda$ reduces the size and
brightness of the higher-order images. As $\theta$ increases, the
shape of the higher-order images changes, and the obscuring effect
of radiation outside the equatorial plane on the dark region inside
the higher-order images becomes more pronounced. However,
differences from isotropic radiation include: (1) At $\theta =
80^\circ$, for isotropic radiation, the higher-order images remain
approximately circular, whereas under anisotropic radiation, they
significantly deform into an ellipsoidal shape. This is because the
intensity of the polar-directed radiation flow surpasses that of the
equatorial radiation flow, causing the intensity distribution of the
higher-order images to be more diffuse and extend farther in the
vertical direction than in the horizontal direction, resulting in an
overall ellipsoidal morphology. (2) Under anisotropic radiation, the
intensity of the direct image is significantly greater than under
isotropic radiation. This is because the fitting function
(\ref{eqI1}) for isotropic radiation gives the average radiation
intensity of fluid elements (emitting particles) in all directions,
while the fitting function (\ref{eqI2}) for anisotropic radiation
gives the radiation intensity of fluid elements in their brightest
direction. Therefore, the direct image intensity is greater under
anisotropic radiation. (3) Observing Fig. (\textbf{i}), it can be
seen that the intensity of the ring structure in the vertical
direction is significantly greater than in the horizontal direction.
This is in stark contrast to the isotropic radiation case, where the
horizontal intensity is markedly greater than the vertical
intensity. Under isotropic radiation, emitting particles radiate
equally in all directions, and the equatorial direction experiences
the maximum degree of Doppler beaming (photons concentrated forward)
and Doppler boosting (photon energy increases), leading to a sharp
enhancement in brightness, thus making the horizontal intensity
greater than the vertical intensity. Under anisotropic radiation,
the emitting particles have their maximum radiation intensity along
the vertical direction, which explains why the vertical intensity
exceeds the horizontal intensity.

The above results indicate that, under the influence of different
radiation types, there are significant differences in the shape and
intensity distribution of higher-order images in shadow images.
These differences become even more pronounced at high observation
inclinations.

\begin{figure}[H]
\centering
\subfigure[$85$~GHz]{\includegraphics[scale=0.5]{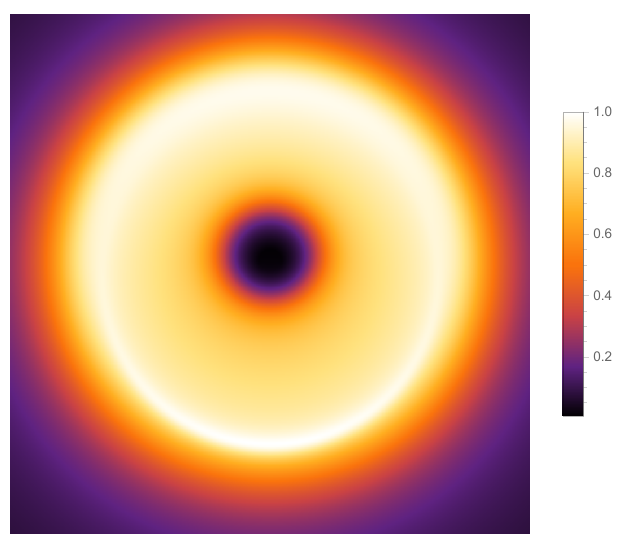}}
\subfigure[$230$~GHz]{\includegraphics[scale=0.5]{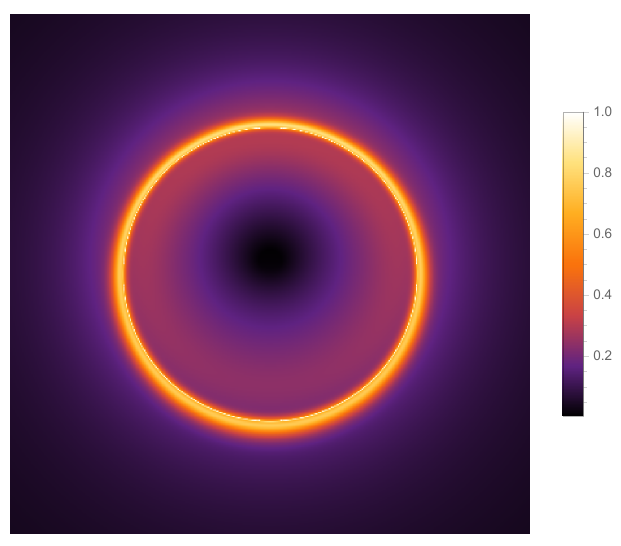}}
\subfigure[$345$~GHz]{\includegraphics[scale=0.5]{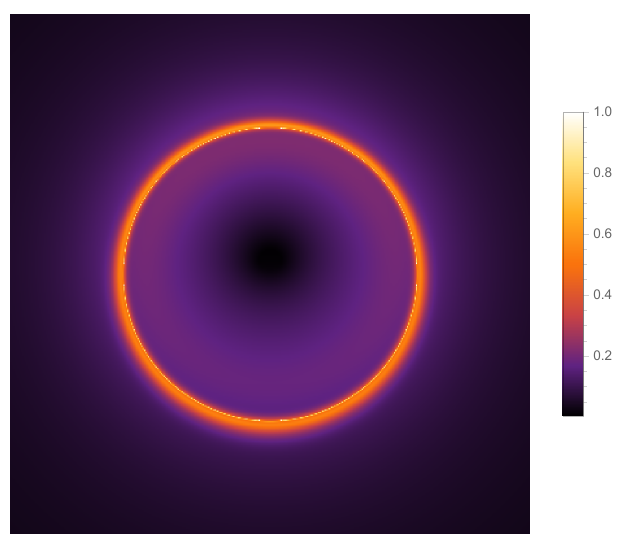}}
\subfigure[Horizontal
direction]{\includegraphics[scale=0.6]{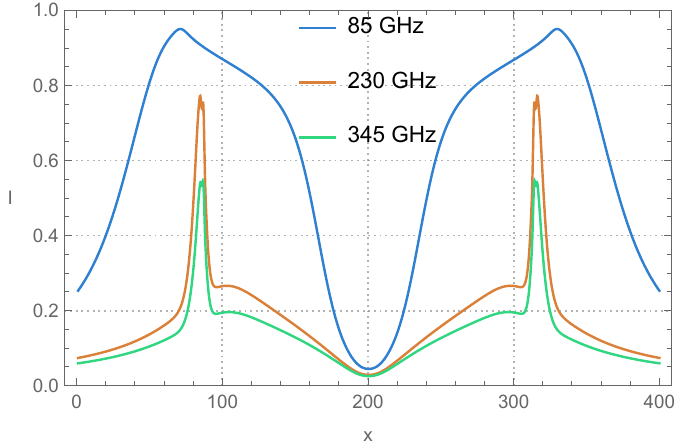}}
\subfigure[Vertical
direction]{\includegraphics[scale=0.6]{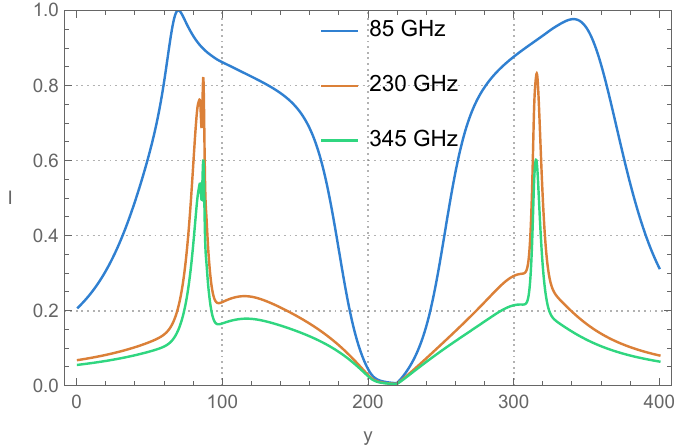}}
\caption{The first row of images shows: the black hole shadow images
under anisotropic radiation for the phenomenological model. The
accretion flow motion mode is infalling motion. From left to right,
the observation frequencies are $85$~GHz, $230$~GHz, and $345$~GHz.
The second row of images shows: the intensity distribution profile
plots at different observation frequencies. The left plot is the
horizontal intensity distribution profile, and the right plot is the
vertical intensity distribution profile. The observer distance is
fixed at $500M$, the field of view is $2^\circ$, the observation
inclination angle is $17^\circ$, and
$\lambda=0.6$.}\label{figyi_curve}
\end{figure}

Figure \textbf{\ref{figyi_curve}} presents the shadow images of the
black hole under anisotropic radiation for the phenomenological
model, along with their characteristics at different observation
frequencies. The first row of images in Fig.
\textbf{\ref{figyi_curve}}, from left to right, corresponds to
observation frequencies of $85$ GHz, $230$ GHz, and $345$ GHz. It
can be observed that when the observation frequency is $85$ GHz, the
distribution range of intensity is the largest. As the observation
frequency increases, the intensity distribution shrinks to a narrow
region, similar to the case of isotropic radiation. However, unlike
isotropic radiation, under anisotropic radiation, the maximum
intensity in the vertical direction is consistently greater than
that in the horizontal direction. This is because, in the isotropic
model, radiation is uniformly distributed, and the difference in
intensity between vertical and horizontal directions is not
significant. In contrast, the anisotropic model introduces
directional dependence, amplifying the radiation efficiency in the
vertical direction, which results in higher maximum intensity in the
vertical direction compared to the horizontal direction.

\subsection{HOU Disk Model}
The HOU disk model was proposed in references \cite{Hou:2023bep,
Zhang:2024lsf}. In this model, the accreting material is confined to
surfaces of constant $\theta$, i.e., $u^\theta \equiv 0$. The
conservation equation for the mass flow can thus be written as:
\begin{equation}
\frac{d}{dr} \left( \sqrt{-g} \rho u^r \right) = 0,
\end{equation}
and its solution is:
\begin{equation}
\rho = \frac{\rho_0}{\sqrt{-g} u^r} \left. \sqrt{-g} u^r
\right|_{r=r_0}.
\end{equation}
Here, $\rho_0 = \rho(r_0)$ is the mass density at the reference
point, typically taken as $r_0 = r_h$. Projecting the conservation
equation of the energy-momentum tensor along $u^\mu$, we obtain:
\begin{equation}
de = \frac{e + p}{\rho} d\rho, \label{eq_de}
\end{equation}
where $e$ is the internal energy of the fluid. Defining $k = T_p /
T_e$ as the proton-to-electron temperature ratio, the internal
energy of the fluid under this approximation satisfies:
\begin{equation}
e = \rho + \rho \frac{3}{2} (k + 2) \frac{m_e}{m_p} \theta_e,
\label{eq_e}
\end{equation}
where $\theta_e = k_B T_e / m_e c^2$ is the dimensionless electron
temperature. Using the ideal gas equation of state, we have:
\begin{equation}
p = n k_B (T_p + T_e) = \rho (1 + k) \frac{m_e}{m_p} \theta_e.
\label{eq_p}
\end{equation}
Substituting Eqs.~(\ref{eq_e}) and (\ref{eq_p}) into
Eq.~(\ref{eq_de}) and integrating, we obtain:
\begin{equation}
\theta_e = (\theta_e)_0 \left( \frac{\rho}{\rho_0}
\right)^{\frac{2(1+k)}{3(2+k)}},
\end{equation}
where $(\theta_e)_0 = \theta_e(r_0)$ is the temperature at a
reference point, again chosen as $r_0 = r_h$. For convenience, we
assume that $\rho(r_h, \theta)$ follows a Gaussian distribution in
the $\theta$ direction, and in the conical solution, we take
$\theta_e(r_h, \theta)$ as a constant:
\begin{align}
\rho(r_h, \theta) &= \rho_h \exp \left[ -\left( \frac{\sin \theta -\sin \theta_J }{\sigma} \right)^2 \right], \\
\theta(r_h, \theta) &= \theta_h.
\end{align}
Here, $\theta_J$ is the average position in the $\theta$ direction,
and $\sigma$ describes the standard deviation of the distribution.
Given the mass density, the number density can be derived from the
relation $\rho = n_e (m_p c^2)$. For M87*, observational results
indicate $\rho_h \approx 1.5 \times 10^3 \, \mathrm{g/cm/s^2}$ and
$\theta_h \approx 16.86$, corresponding to $n_h = 10^6 \,
\mathrm{cm^{-3}}$ and $T_h = 10^{11} \, \mathrm{K}$. In a
spherically symmetric spacetime, the magnetic field configuration
simplifies to:
\begin{equation}
B^\mu = \frac{\Xi}{\sqrt{-g} u^r} (u_t u^\mu + \delta^\mu_t),
\end{equation}
where $\Xi = F_{\theta\phi}$ is a component of the electromagnetic
tensor. Note that $u^r$ appears in the denominator, so orbital
motion cannot be adopted for the fluid in the HOU disk model. For
the above expression, we use the split monopole solution:
\begin{equation}
\Xi = \Xi_0 \, \mathrm{sign}(\cos \theta) \sin \theta.
\end{equation}
This analysis indicates that the HOU disk model must account for the
magnetic field direction; thus, the electron radiation model employs
the anisotropic radiation formula~(\ref{eqI2}). For the HOU disk
model, the fluid motion can be chosen as either free-fall motion or
a combination of free-fall and circular motion. One may also
consider the ballistic approximation, where the fluid moves along
geodesics (for related calculations in Kerr spacetime, refer to
\cite{Hou:2023bep}). However, purely circular motion cannot be
adopted, as the fluid in the HOU disk model requires a radial
velocity; otherwise, the magnetic field would diverge. This section
also considers synchrotron radiation. Therefore, given the magnetic
field, number density, and temperature, we can similarly calculate
the intensity based on Eqs.~(\ref{eq_inu}), (\ref{eq_jnu}), and
(\ref{eq_alphanu_bnu}). Below, we will present the numerical results
for the Hou disk model.

\begin{figure}[H]
\centering
\subfigure[$\lambda=0.1,\theta=17^\circ$]{\includegraphics[scale=0.37]{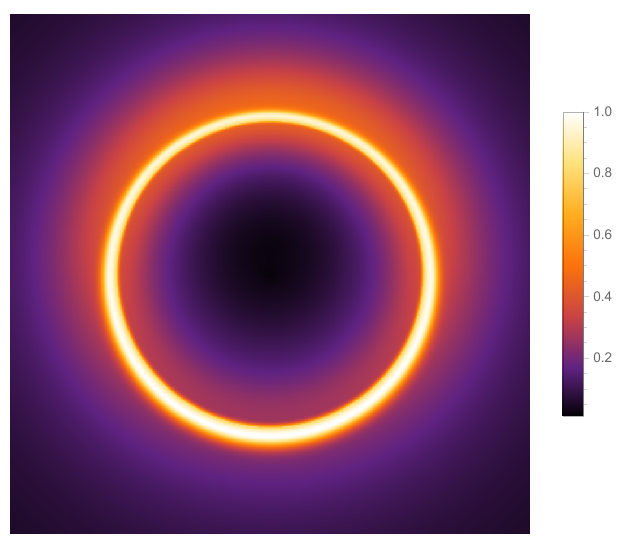}}
\subfigure[$\lambda=0.3,\theta=17^\circ$]{\includegraphics[scale=0.37]{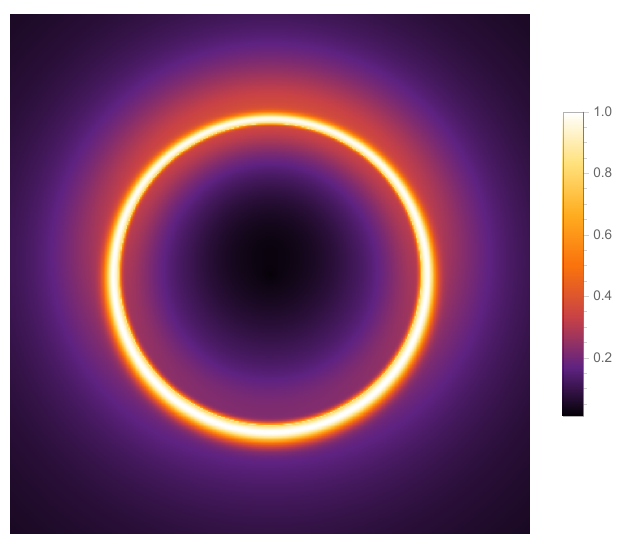}}
\subfigure[$\lambda=0.6,\theta=17^\circ$]{\includegraphics[scale=0.37]{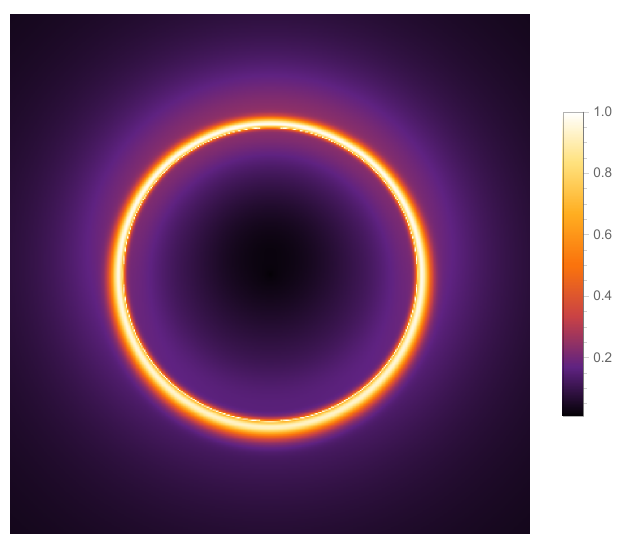}}
\subfigure[$\lambda=0.9,\theta=17^\circ$]{\includegraphics[scale=0.37]{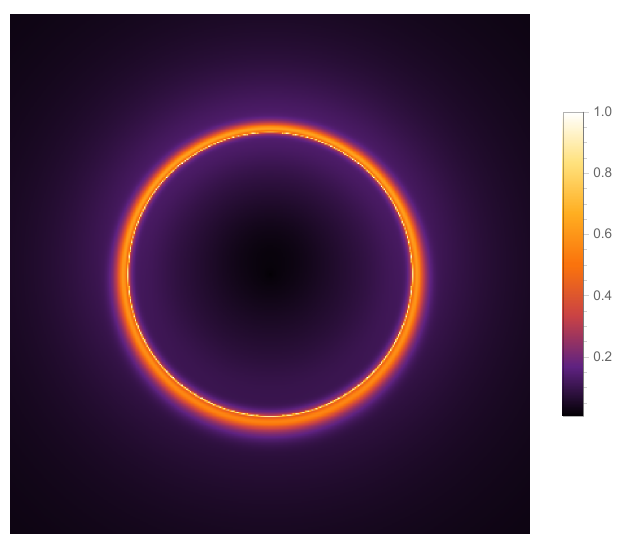}}
\subfigure[$\lambda=0.1,\theta=50^\circ$]{\includegraphics[scale=0.37]{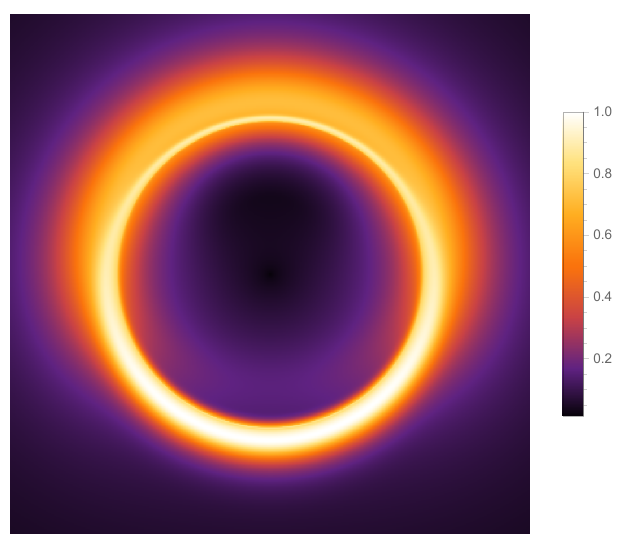}}
\subfigure[$\lambda=0.3,\theta=50^\circ$]{\includegraphics[scale=0.37]{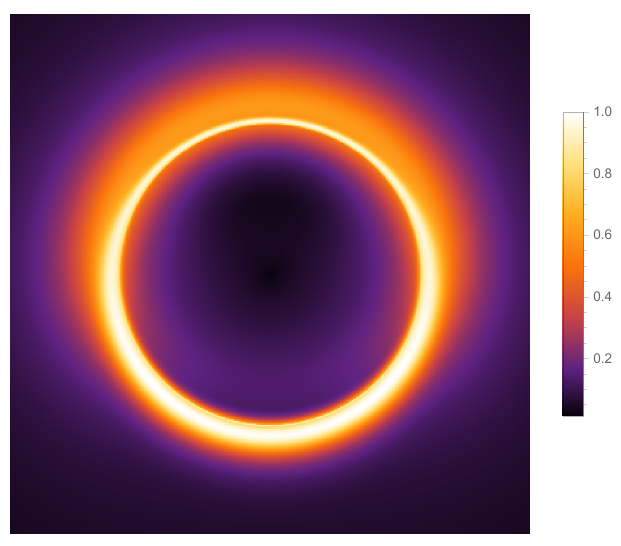}}
\subfigure[$\lambda=0.6,\theta=50^\circ$]{\includegraphics[scale=0.37]{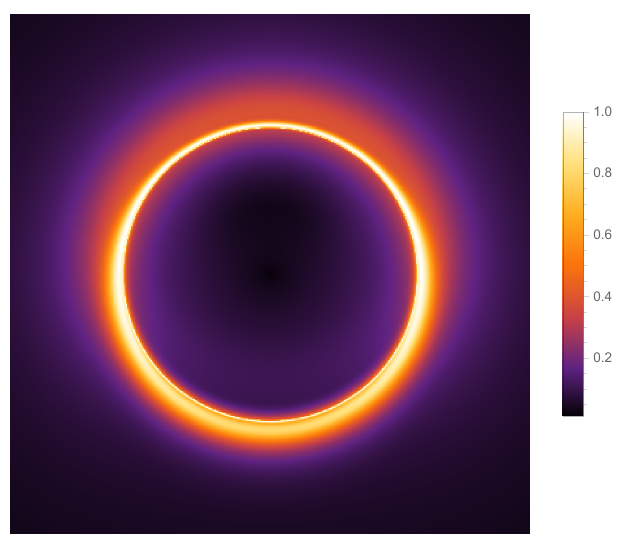}}
\subfigure[$\lambda=0.9,\theta=50^\circ$]{\includegraphics[scale=0.37]{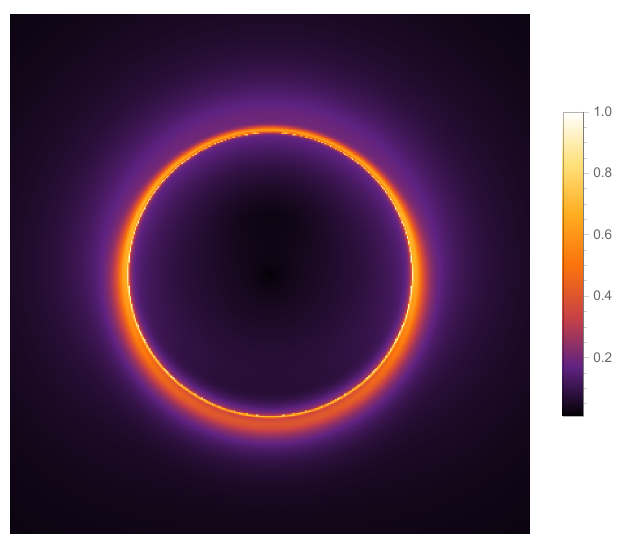}}
\subfigure[$\lambda=0.1,\theta=80^\circ$]{\includegraphics[scale=0.37]{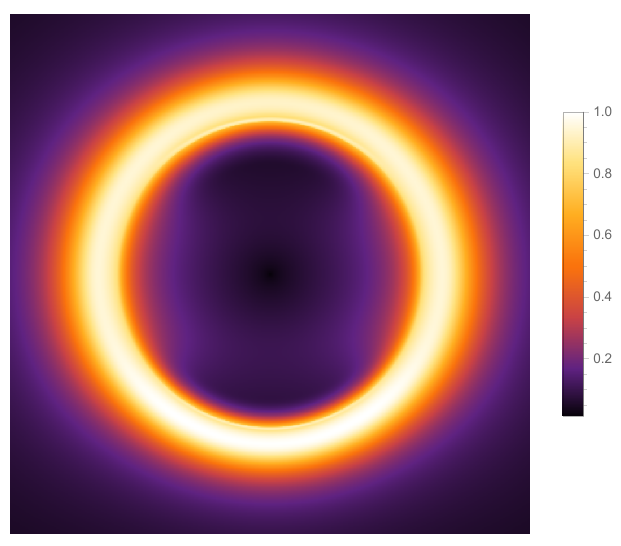}}
\subfigure[$\lambda=0.3,\theta=80^\circ$]{\includegraphics[scale=0.37]{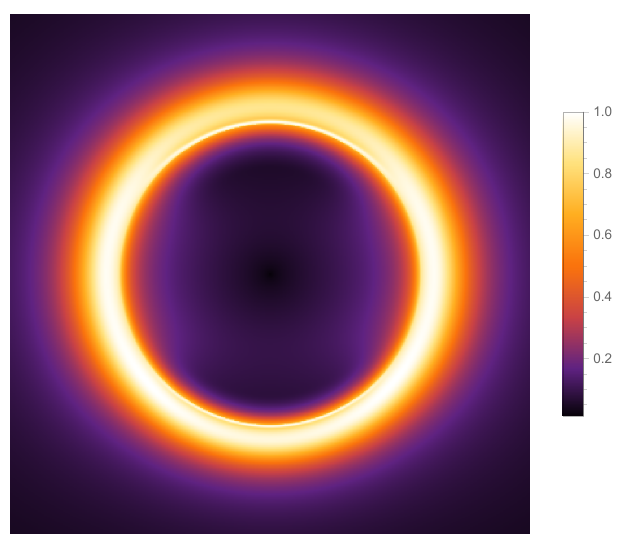}}
\subfigure[$\lambda=0.6,\theta=80^\circ$]{\includegraphics[scale=0.37]{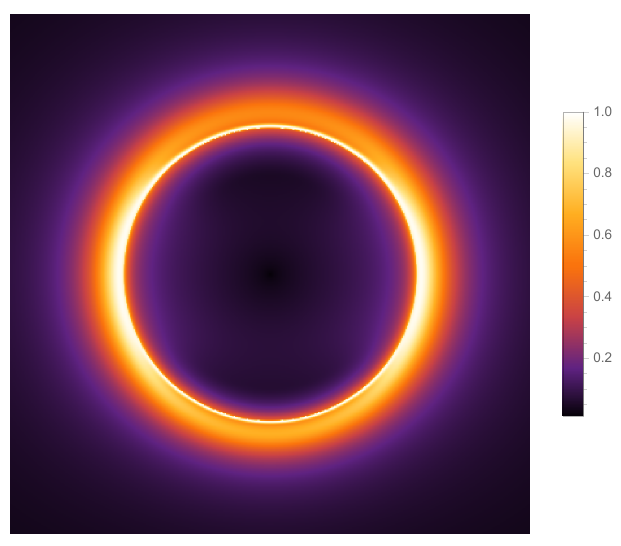}}
\subfigure[$\lambda=0.9,\theta=80^\circ$]{\includegraphics[scale=0.37]{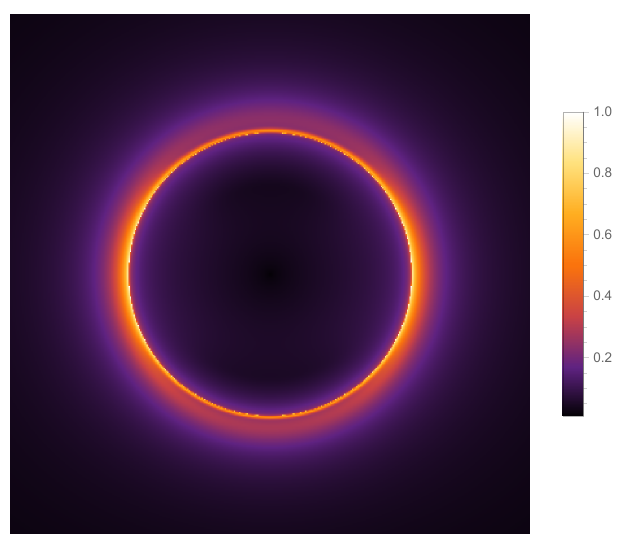}}
\caption{Black hole shadow images under the Hou disk model. The
accretion flow motion mode is infalling motion. From left to right,
the parameter $\lambda$ takes values of $0.1$, $0.3$, $0.6$, and
$0.9$, respectively. From top to bottom, the observation inclination
$\theta$ takes values of $17^\circ$, $50^\circ$, and $80^\circ$,
respectively. The observer distance is fixed at $500M$, the field of
view is $2^\circ$, and the observation frequency is $230$
GHz.}\label{fighou}
\end{figure}

Figure \textbf{\ref{fighou}} shows the black hole shadow images for
the Hou disk model under anisotropic radiation, with an observation
frequency of $230$ GHz and the accretion flow in infalling motion.
The bright ring in the images still represents the higher-order
images, and the dark region inside the higher-order images
originates from the event horizon. Increasing $\lambda$ reduces the
size of the higher-order images. Increasing $\theta$ enhances the
brightness of the direct images outside the higher-order images but
hardly changes the size of the higher-order images, which is in
sharp contrast to the phenomenological model. Meanwhile, for the Hou
disk model, the obscuration effect of out of equatorial plane
radiation on the event horizon contour is weakened at high
observation inclinations.

\begin{figure}[H]
\centering \subfigure[Isotropic
radiation]{\includegraphics[scale=0.5]{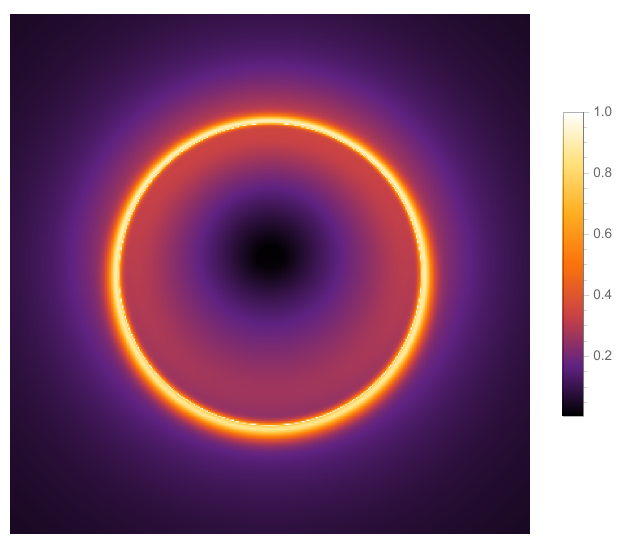}}
\subfigure[Isotropic
radiation]{\includegraphics[scale=0.5]{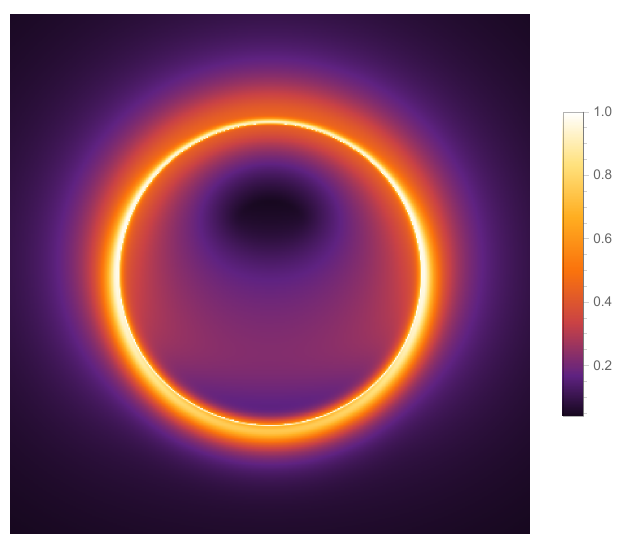}}
\subfigure[Isotropic
radiation]{\includegraphics[scale=0.5]{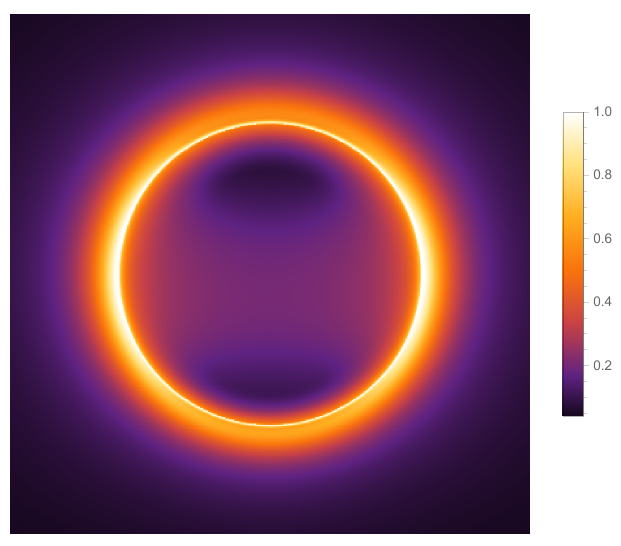}}
\subfigure[Anisotropic
radiation]{\includegraphics[scale=0.5]{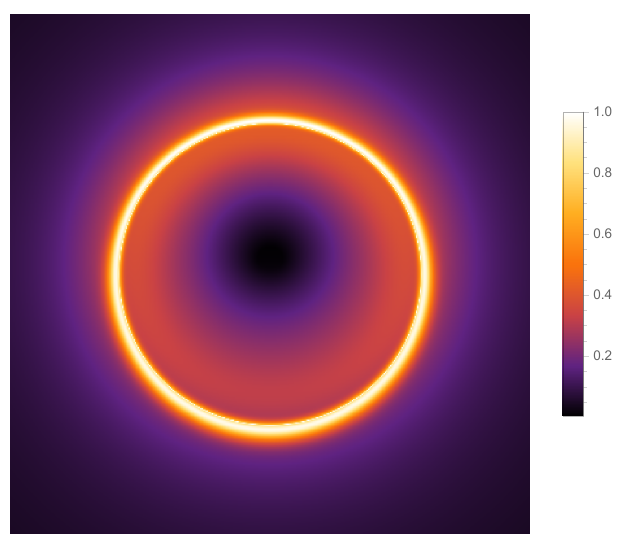}}
\subfigure[Anisotropic
radiation]{\includegraphics[scale=0.5]{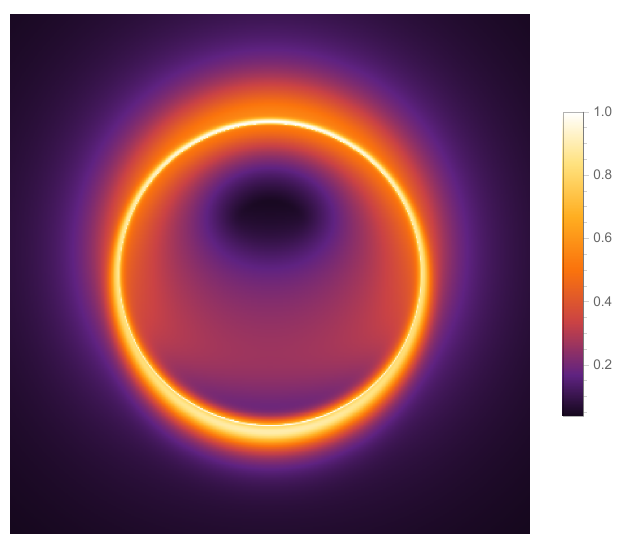}}
\subfigure[Anisotropic
radiation]{\includegraphics[scale=0.5]{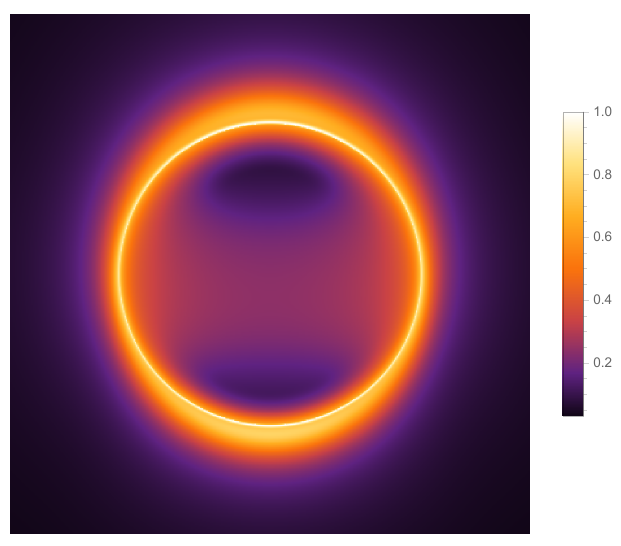}}
\subfigure[Hou disk
model]{\includegraphics[scale=0.5]{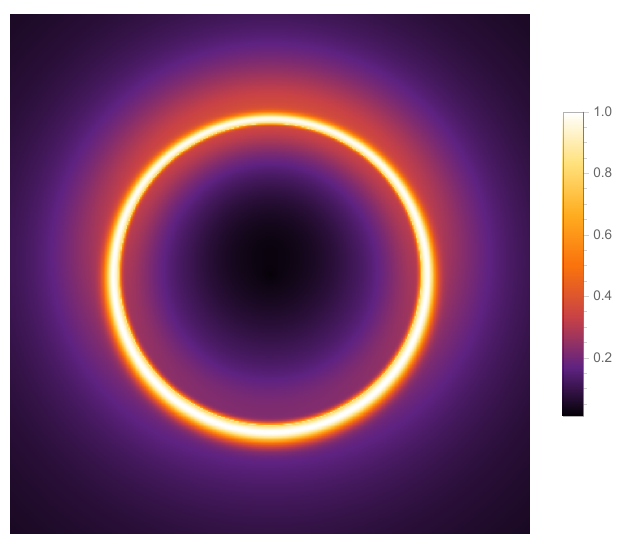}}
\subfigure[Hou disk
model]{\includegraphics[scale=0.5]{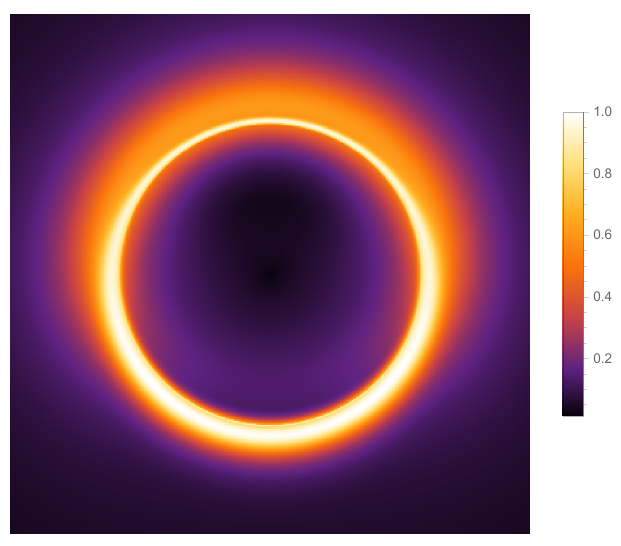}}
\subfigure[Hou disk
model]{\includegraphics[scale=0.5]{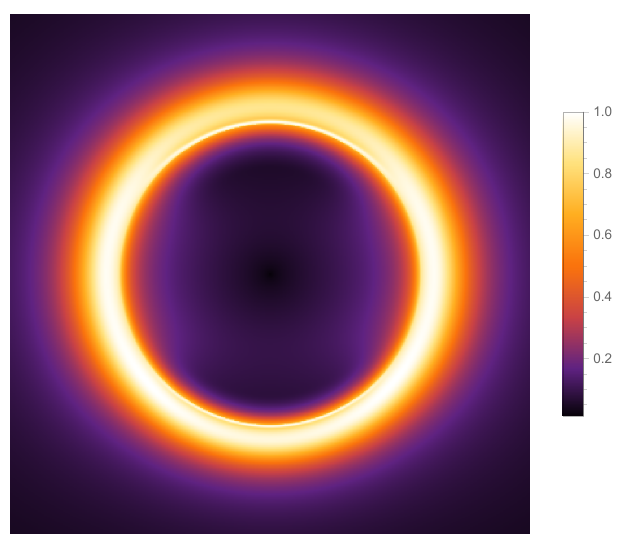}}
\caption{Influence of accretion disk models on black hole shadows at
different observation inclinations. The first and second rows
correspond to the phenomenological model, while the third row
corresponds to the Hou disk model. From left to right, the images
correspond to observation inclinations $\theta = 17^\circ$,
$50^\circ$, and $80^\circ$, respectively. The accretion flow motion
mode is infalling motion, with the parameter $\lambda$ is fixed at
$0.3$ and the observation frequency at $230$ GHz.}\label{figcompare}
\end{figure}

\begin{figure}[H]
\centering
\subfigure[$85$~GHz]{\includegraphics[scale=0.5]{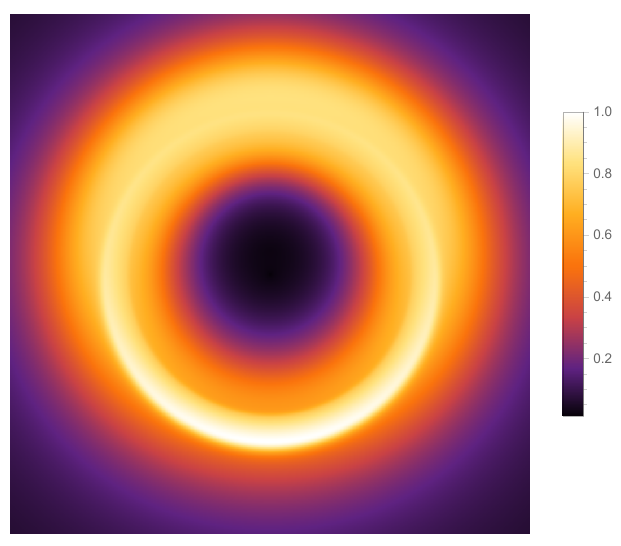}}
\subfigure[$230$~GHz]{\includegraphics[scale=0.5]{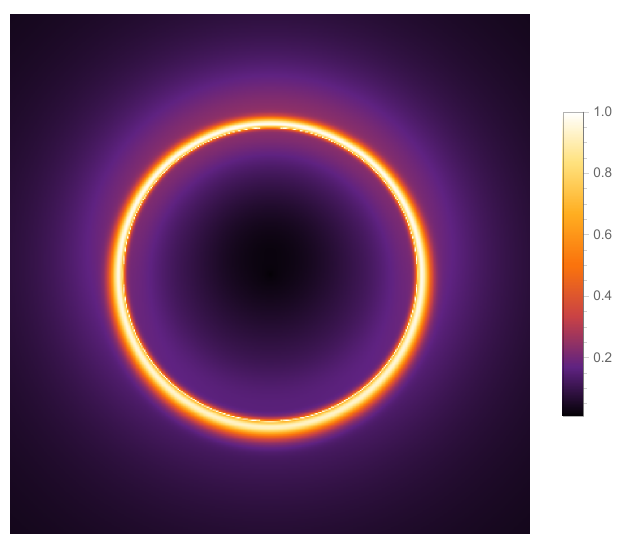}}
\subfigure[$345$~GHz]{\includegraphics[scale=0.5]{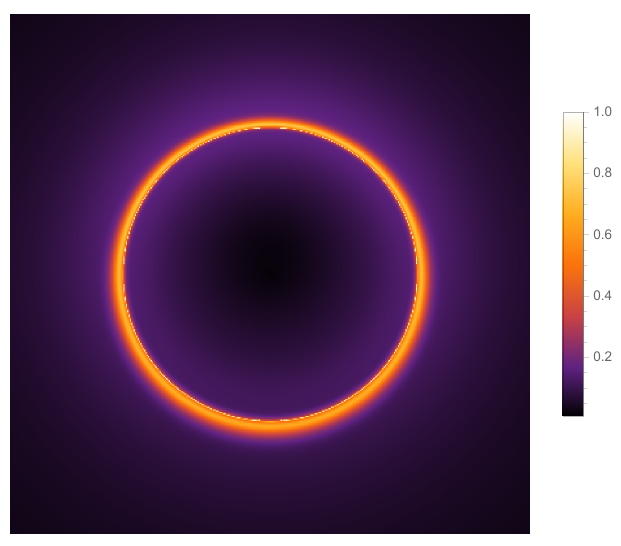}}
\subfigure[Horizontal
direction]{\includegraphics[scale=0.6]{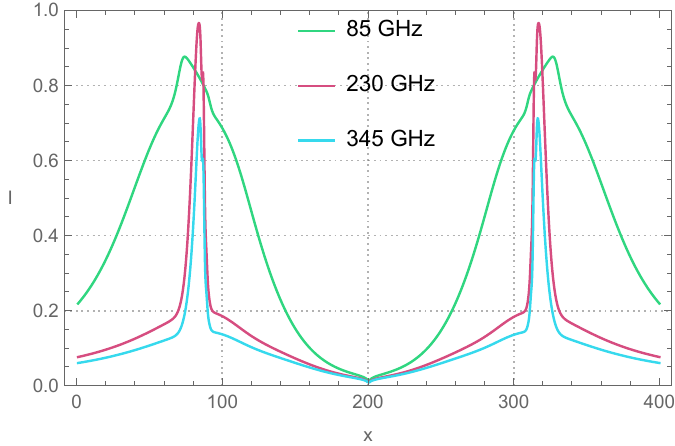}}
\subfigure[Vertical
direction]{\includegraphics[scale=0.6]{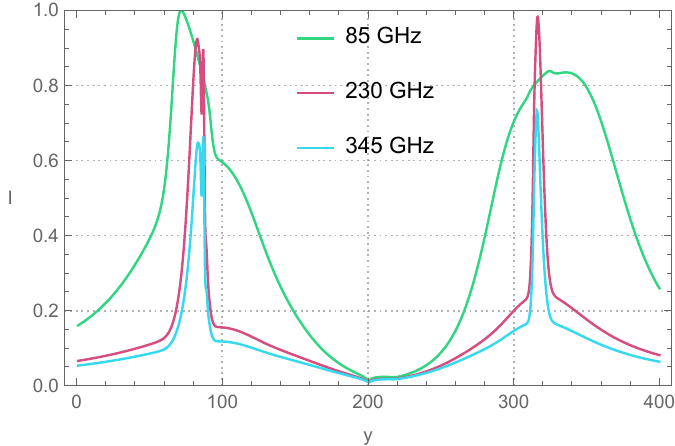}}
\caption{The first row of images shows the black hole shadow images
under the Hou disk model. The accretion flow motion mode is
infalling motion. From left to right, the observation frequencies
are $85$ GHz, $230$ GHz, and $345$ GHz, respectively. The second row
of images shows the intensity distribution profiles at different
observation frequencies. The left image is the horizontal intensity
distribution profile, and the right image is the vertical intensity
distribution profile. The observer distance is fixed at $500M$, the
field of view is $2^\circ$, the observation inclination is
$17^\circ$, and $\lambda=0.6$.}\label{fighou_curve}
\end{figure}

To intuitively compare the differences between the two models
mentioned above, Fig. \textbf{\ref{figcompare}} illustrates the
influence of the accretion disk model on the black hole shadow at
different observer inclinations. It can be observed that at the same
observation inclination, the direct image in the phenomenological
model is significantly brighter than that in the Hou disk model.
Additionally, at high observer inclinations ($\theta = 80^\circ$
(third column)), the obscuration effect of out of equatorial plane
radiation on the event horizon contour is more pronounced in the
phenomenological model, indicating that the gravitational lensing
effect is stronger in the phenomenological model. Figure
\textbf{\ref{fighou_curve}} displays the variation of the black hole
shadow image with observation frequency for the Hou accretion disk
model. The first row of images in Fig. \textbf{\ref{fighou_curve}},
from left to right, corresponds to observation frequencies of $85$
GHz, $230$ GHz, and $345$ GHz. It can be observed that, similar to
the previous discussion, when the observation frequency is $85$ GHz,
the intensity distribution range is the largest. Each curve exhibits
two significant peaks. The ring-like structure corresponding to the
maximum intensity represents the higher-order images of the
accretion disk, while the region outside the peaks corresponds to
the direct images of the accretion disk.

\section{Polarization Imaging}
For polarization imaging, only the anisotropic radiation under the
Hou disk model is considered. As in the previous section, the
accretion flow motion mode is chosen as infalling motion. Under the
WKB approximation, light propagation satisfies the radiative
transfer equation:
\begin{equation}
k^\mu \nabla_\mu S^{\alpha\beta} = J^{\alpha\beta} +
H^{\alpha\beta\mu\nu} S_{\mu\nu}, \label{eq39}
\end{equation}
where $k^\mu$ is the wave vector of the light ray, $S^{\alpha\beta}$
is the polarization tensor used to characterize the polarization
state of the light, $J^{\alpha\beta}$ describes the emission
properties of the radiation source, and $H^{\alpha\beta\mu\nu}$
characterizes the response of the propagation medium to the light
ray, typically including absorption effects and Faraday rotation
effects. Numerical solutions to the radiative transfer equation can
refer to the open-source code Coport 1.0 \cite{Huang:2024bar}.
Building on Coport 1.0, we can simplify the calculations using the
gauge invariance of $S^{\alpha\beta}$ within a simple
parallel-transported frame. In this case, the covariant radiative
transfer equation (\ref{eq39}) is decomposed into two parts. The
first part accounts for gravitational effects:
\begin{equation}
k^\mu \nabla_\mu f^a = 0, \quad f^a k_a = 0,
\end{equation}
where $f^\mu$ is a normalized spacelike vector orthogonal to
$k^\mu$. The second part is the radiative transfer equation:
\begin{equation}
\frac{d}{d\lambda} \mathbf{S} = \mathbf{R}(\chi) \mathbf{J} -
\mathbf{R}(\chi) \mathbf{M} \mathbf{R}(-\chi) \mathbf{S},
\end{equation}
where
\begin{equation}
\mathbf{S} =
\begin{pmatrix}
\mathcal{I} \\
\mathcal{Q} \\
\mathcal{U}  \\
\mathcal{V}
\end{pmatrix}, \quad
\mathbf{J} = \frac{1}{\nu^2}
\begin{pmatrix}
j_I \\
j_Q \\
j_U \\
j_V
\end{pmatrix}, \quad
\mathbf{M} = \nu
\begin{pmatrix}
a_I & a_Q & a_U & a_V \\
a_Q & a_I & r_V & -r_U \\
a_U & -r_V & a_I & r_Q \\
a_V & r_U & -r_Q & a_I
\end{pmatrix}.
\end{equation}

$\mathbf{R}(\chi)$ is the rotation matrix used to rotate between the
synchrotron emission basis and the parallel-transported reference
basis:
\begin{equation}
\mathbf{R}(\chi) =
\begin{pmatrix}
1 & 0 & 0 & 0 \\
0 & \cos(2\chi) & -\sin(2\chi) & 0 \\
0 & \sin(2\chi) & \cos(2\chi) & 0 \\
0 & 0 & 0 & 1
\end{pmatrix}. \label{eq43}
\end{equation}
Here, the rotation angle $\chi$ is the angle between the reference
vector $f^\mu$ and the local magnetic field $b^\mu$ in the
transverse subspace of the light ray:
\begin{equation}
\chi = \mathrm{sign}(\epsilon_{\mu\nu\alpha\beta} u^\mu f^\nu b^\rho
k^\sigma) \arccos\left( \frac{P^{\mu\nu} f_\mu
b_\nu}{\sqrt{(P^{\mu\nu} f_\mu f_\nu)(P^{\alpha\beta} b_\alpha
b_\beta)}} \right),
\end{equation}
where $P^{\mu\nu}$ is the induced metric in the transverse subspace.
At the observer's location, the Stokes parameters need to be
projected onto the observer's screen using the rotation matrix
(\ref{eq43}), where the rotation angle is:
\begin{equation}
\chi_o = \mathrm{sign}(\epsilon_{\mu\nu\rho\sigma} u^\mu f^\nu
d^\rho k^\sigma) \arccos\left( \frac{P^{\mu\nu} f_\mu
d_\nu}{\sqrt{(P^{\mu\nu} f_\mu f_\nu) (P^{\alpha\beta} d_\alpha
d_\beta)}} \right),
\end{equation}
where $d^\mu$ is the $y$-axis direction of the screen, chosen in
this paper as $d^\mu = -\partial_\theta^\mu$. The projection results
are:
\begin{equation}
\mathcal{I_o} = \mathcal{I}, \quad \mathcal{Q_o} = \mathcal{Q} \cos
\chi_o - \mathcal{U} \sin \chi_o, \quad \mathcal{U_o} = \mathcal{Q}
\sin \chi_o + \mathcal{U} \cos \chi_o, \quad \mathcal{V_o} =
\mathcal{V}.
\end{equation}

The Stokes parameter $\mathcal{I_o}$ reflects the intensity of the
light. $\mathcal{V_o}$ reflects information about circular
polarization: if $\mathcal{V_o}$ is positive, it corresponds to
left-handed circular polarization; otherwise, it corresponds to
right-handed circular polarization. $\mathcal{Q_o}$ and
$\mathcal{U_o}$ reflect information about the electric field
$\vec{E} = (E_x, E_y)$:
\begin{equation}
\mathcal{Q_o} = E_x^2 - E_y^2, \quad \mathcal{U_o} = 2E_x E_y.
\end{equation}
If $\mathcal{U_o}$ is positive, $E_x$ and $E_y$ have the same sign,
and $\vec{E}$ lies in the first or third quadrant; if
$\mathcal{U_o}$ is negative, $E_x$ and $E_y$ have opposite signs,
and $\vec{E}$ lies in the second or fourth quadrant. The sign of
$\mathcal{Q_o}$ reflects the relationship between $\vec{E}$ and the
lines $y = x$ or $y = -x$. On the projection screen, the
polarization intensity corresponding to the polarization vector
$\vec{f}$ and the electric vector position angle (EVPA) are
calculated as:
\begin{equation}
P_o = \sqrt{\mathcal{Q_o}^2 + \mathcal{U_o}^2}, \quad
\Phi_{\mathrm{EVPA}} = \frac{1}{2} \arctan\left(
\frac{\mathcal{U_o}}{\mathcal{Q_o}} \right). \label{eq48}
\end{equation}

\begin{figure*}[htbp]
  \centering
  \includegraphics[width=0.45\textwidth]{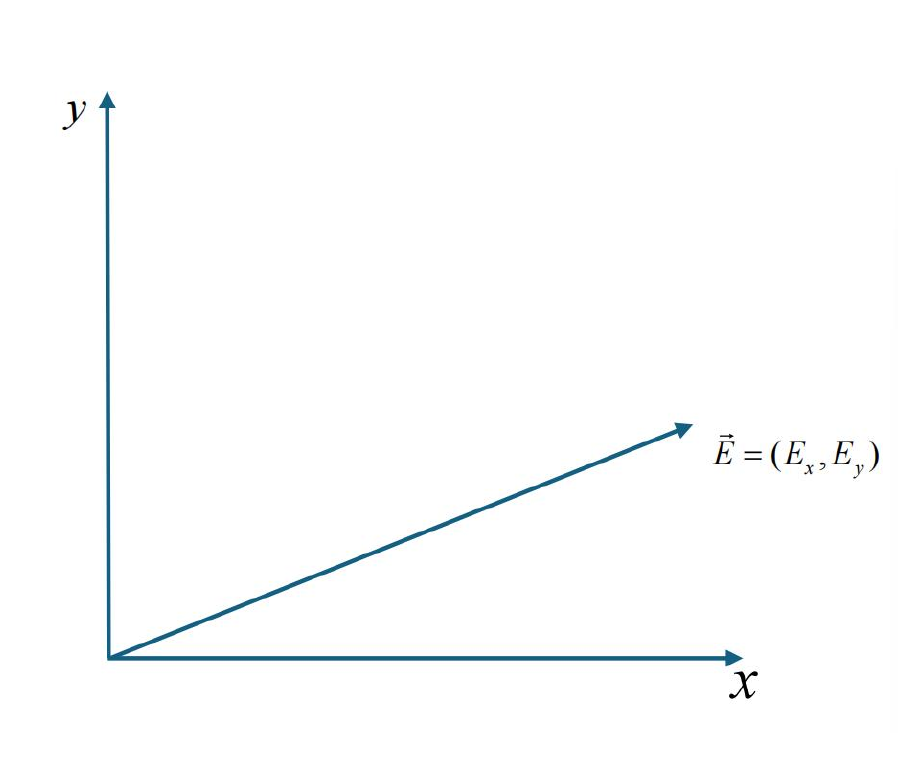}
  \caption{For the $\vec{E}$ shown in the figure, $\mathcal{Q_o} > 0$, $\mathcal{U_o} > 0$.}
  \label{fig1}
\end{figure*}

\section{Numerical Results}
\subsection{Polarization Images Under the HOU Disk Model}
\begin{figure}[H]
\centering \subfigure[Stokes
$~~\mathcal{I_o}$]{\includegraphics[scale=0.6]{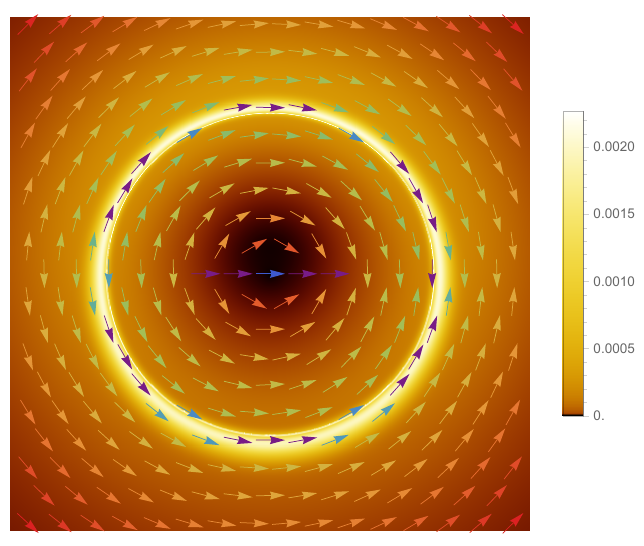}}
\subfigure[Stokes
$~~\mathcal{Q_o}$]{\includegraphics[scale=0.6]{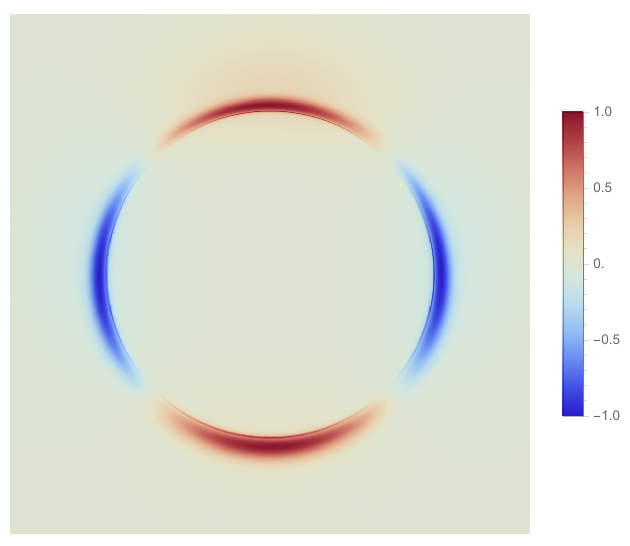}}
\subfigure[Stokes
$~~\mathcal{U_o}$]{\includegraphics[scale=0.6]{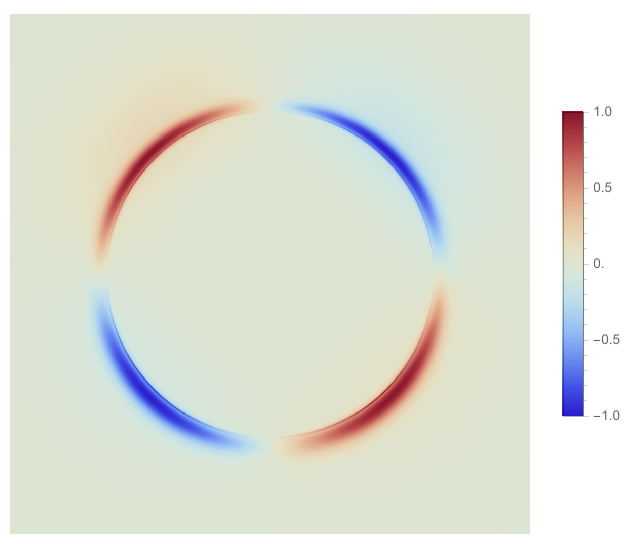}}
\subfigure[Stokes
$~~\mathcal{V_o}$]{\includegraphics[scale=0.6]{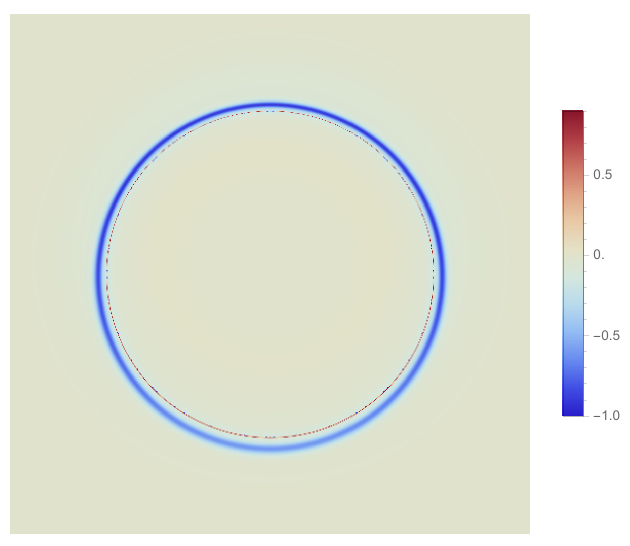}}
\caption{Computation results of Stokes parameters $\mathcal{I_o}$,
$\mathcal{Q_o}$, $\mathcal{U_o}$, and $\mathcal{V_o}$ under the Hou
disk model. The accretion flow motion mode is infalling motion, with
fixed parameters $\lambda=0.6$, $\theta=17^\circ$, and an
observation frequency of 230 GHz.}\label{figIQUV}
\end{figure}

Figure \textbf{\ref{figIQUV}} presents the numerical results for the
Stokes parameters $\mathcal{I_o}$, $\mathcal{Q_o}$, $\mathcal{U_o}$,
and $\mathcal{V_o}$. The accretion flow motion mode is infalling
motion, with fixed parameters $\lambda = 0.6$ and $\theta =
17^\circ$, and an observation frequency of 230 GHz. Here,
$\mathcal{I_o}$ reflects the intensity distribution. The arrows
represent the polarization vectors $\vec{f}$ calculated using Eq.
(\ref{eq48}), where their length and color depth indicate the
polarization intensity $P_o$, and their direction corresponds to the
electric vector position angle $\Phi_{\mathrm{EVPA}}$. Since
$\vec{f}$ is always perpendicular to the magnetic field $\vec{b}$,
it can be inferred that the magnetic field is approximately radially
distributed. Combining $\mathcal{Q_o}$ and $\mathcal{U_o}$ allows
for a qualitative determination of the direction of the electric
vector $\vec{E}$, while $\mathcal{V_o} < 0$ indicates right-handed
circular polarization. The parameters $\mathcal{Q_o}$,
$\mathcal{U_o}$, and $\mathcal{V_o}$ peak in the region of the
higher-order images and rapidly decay to zero away from this region.
Given that $\mathcal{I_o}$ already contains the complete information
of the polarization image, only $\mathcal{I_o}$ will be analyzed in
the subsequent discussion.

\begin{figure}[H]
\centering
\subfigure[$\lambda=0.01,\theta=17^\circ$]{\includegraphics[scale=0.37]{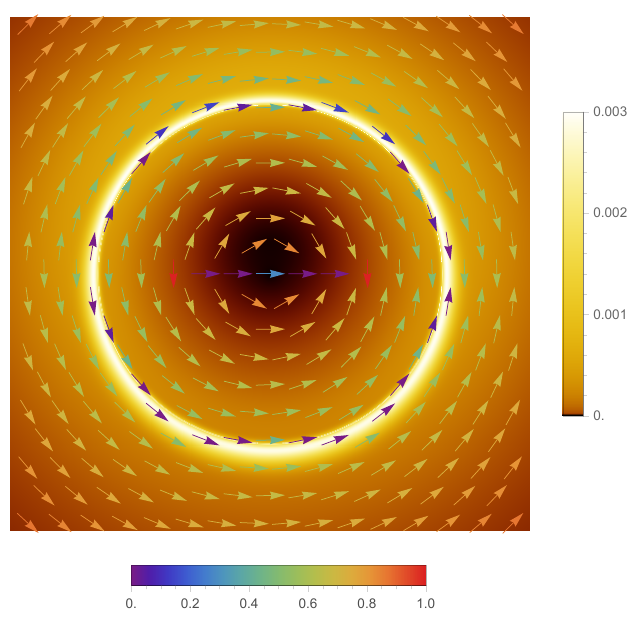}}
\subfigure[$\lambda=0.3,\theta=17^\circ$]{\includegraphics[scale=0.37]{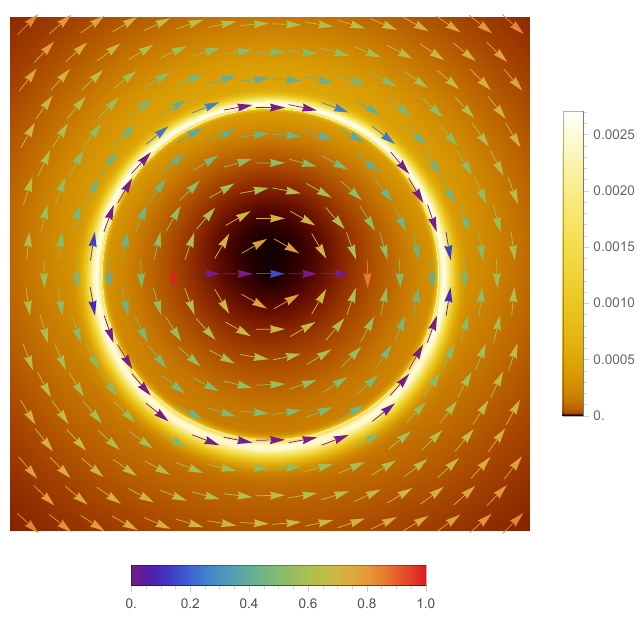}}
\subfigure[$\lambda=0.6,\theta=17^\circ$]{\includegraphics[scale=0.37]{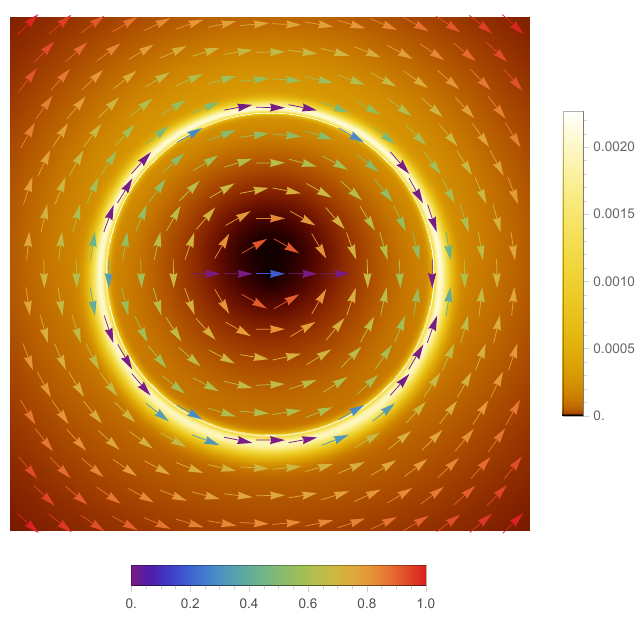}}
\subfigure[$\lambda=0.99,\theta=17^\circ$]{\includegraphics[scale=0.37]{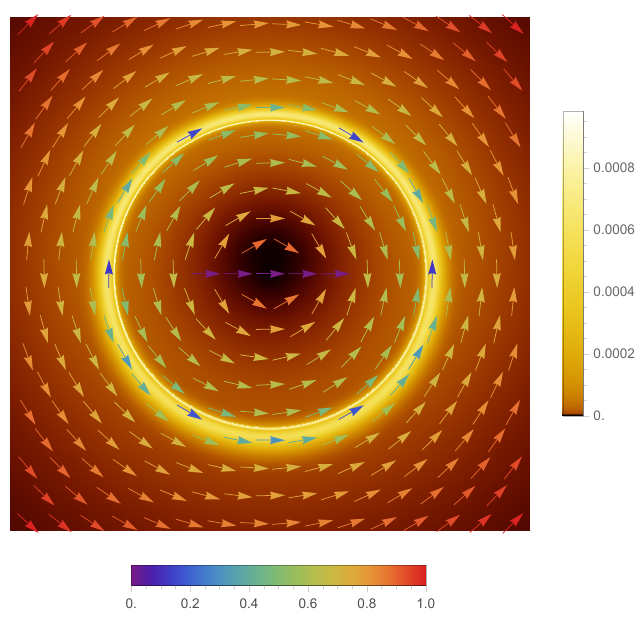}}
\subfigure[$\lambda=0.01,\theta=50^\circ$]{\includegraphics[scale=0.37]{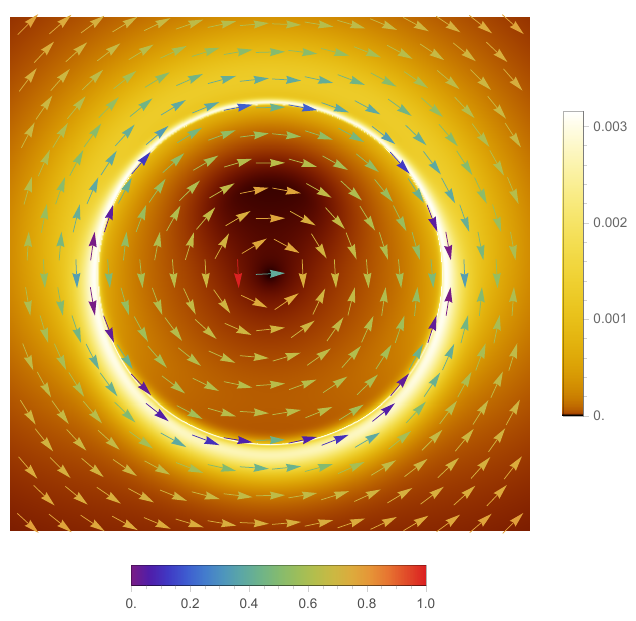}}
\subfigure[$\lambda=0.3,\theta=50^\circ$]{\includegraphics[scale=0.37]{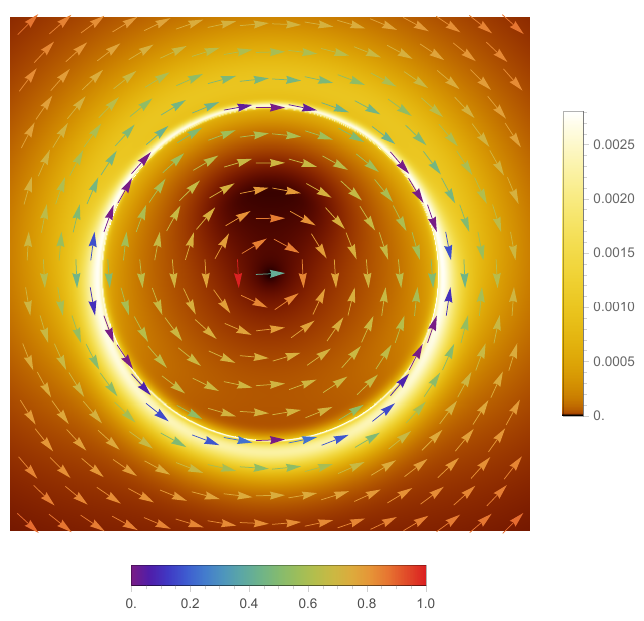}}
\subfigure[$\lambda=0.6,\theta=50^\circ$]{\includegraphics[scale=0.37]{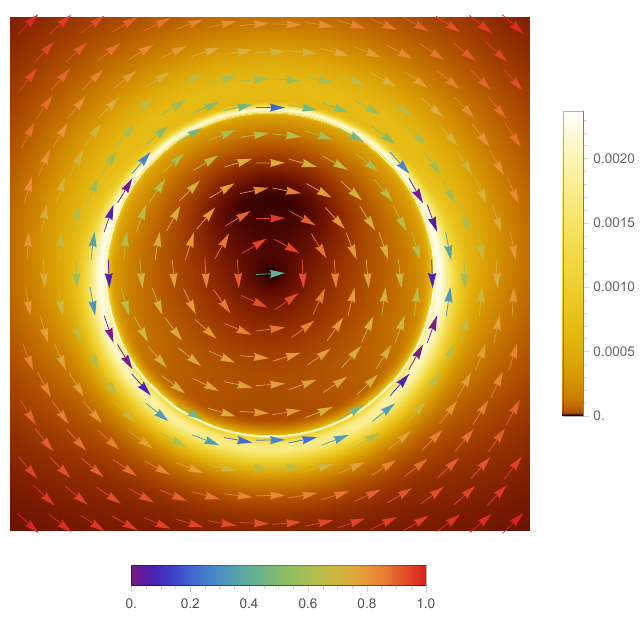}}
\subfigure[$\lambda=0.99,\theta=50^\circ$]{\includegraphics[scale=0.37]{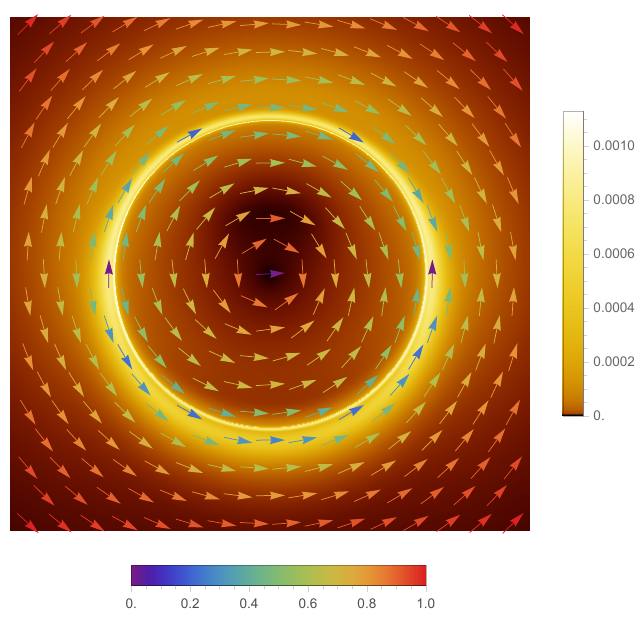}}
\subfigure[$\lambda=0.01,\theta=80^\circ$]{\includegraphics[scale=0.37]{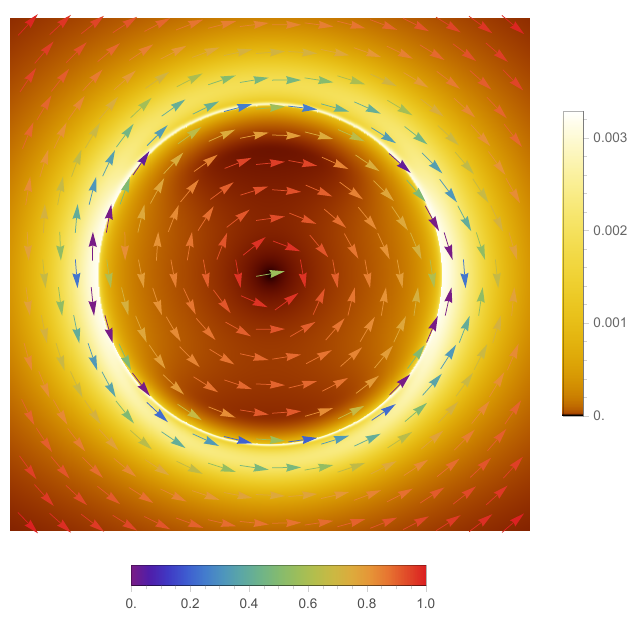}}
\subfigure[$\lambda=0.3,\theta=80^\circ$]{\includegraphics[scale=0.37]{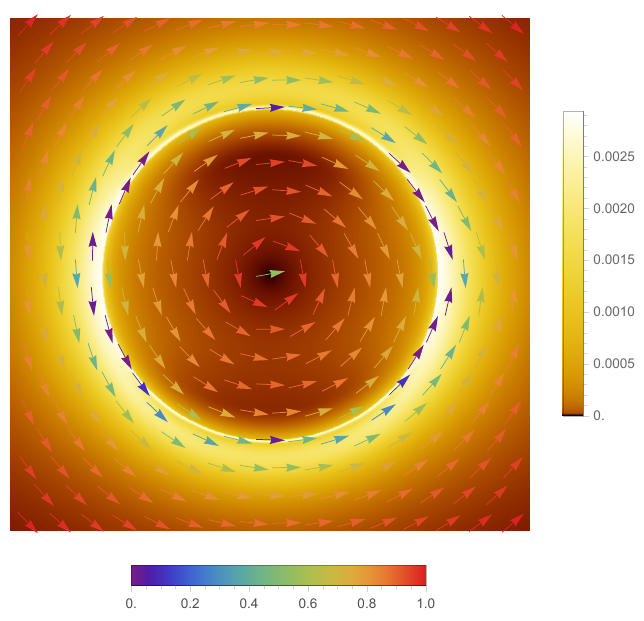}}
\subfigure[$\lambda=0.6,\theta=80^\circ$]{\includegraphics[scale=0.37]{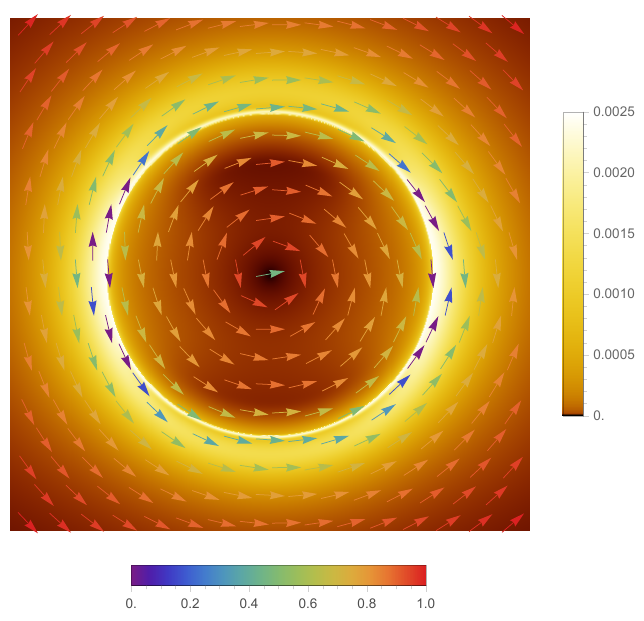}}
\subfigure[$\lambda=0.99,\theta=80^\circ$]{\includegraphics[scale=0.37]{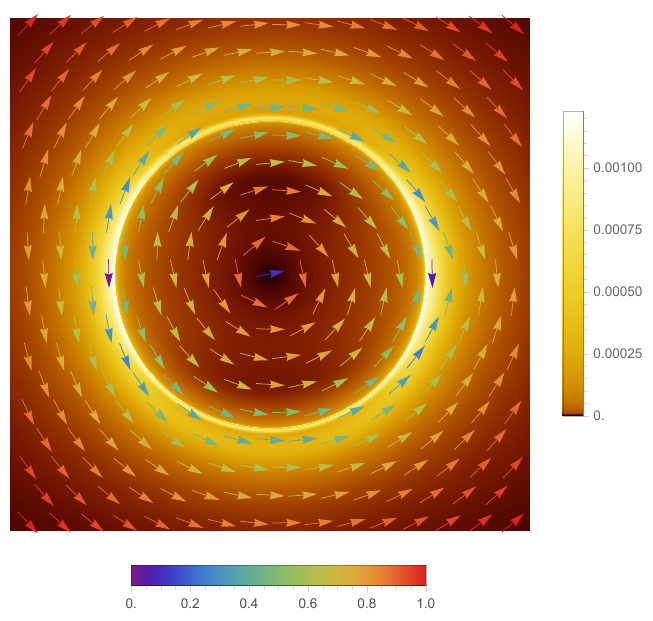}}
\caption{Polarization images of the black hole shadow under the Hou
disk model. The accretion flow motion mode is infalling motion. From
left to right, the parameter $\lambda$ takes values of $0.01$,
$0.3$, $0.6$, and $0.99$, respectively. From top to bottom, the
observation inclination $\theta$ takes values of $17^\circ$,
$50^\circ$, and $80^\circ$, respectively. The observer distance is
fixed at $600M$, the field of view is $1.5^\circ$, and the
observation frequency is $230$ GHz.}\label{figpolar}
\end{figure}

Figure \textbf{\ref{figpolar}} illustrates the influence of the
parameter $\lambda$ on the polarization images under the Hou disk
model. The bright rings in the images correspond to the higher-order
images, and the dark region inside originates from the event
horizon. The images show that the polarization intensity $P_o$ in
the region of the higher-order images is significantly higher than
in other regions and rapidly weakens away from this area. The
results indicate that $\lambda$ determines the intrinsic structure
of spacetime, while $\theta$ depends on the observer's orientation,
and together they influence the polarization characteristics. It is
particularly important to note that in the thin disk model,
radiation cannot escape beyond the event horizon, so polarization
effects cannot be observed in the inner shadow region. In contrast,
in the thick disk model, gravitational lensing causes the event
horizon contour to be obscured by radiation from outside the
equatorial plane, leading to the presence of polarization vectors
across the entire imaging plane.

\section{Conclusion}
In this work, we investigate the visual and polarization
characteristics of a spherically symmetric GB black hole within the
mechanism of various thick disk models. Initially, we defined the
background of the GB black hole and defined null
geodesics and photon sphere. Consequently, we investigated two
representative models of geometrically thick accretion flows: a
phenomenological RIAF model and an analytical BAAF model. The
synchrotron emission produced by thermal electrons in the magnetized
plasma was calculated by numerically integrating the null geodesic
equations with the GRRT approach, yielding the corresponding black
hole images.

In the case of the RIAF model, we have investigated both the isotropic
and anisotropic radiation scenarios with the infalling motion mode
for different values of GB coupling parameter $\lambda$, observer's
positions and observed frequencies. The obtained results show that,
in isotropic radiation scenarios, increasing $\lambda$ reduces both
the size and brightness of the higher-order image, while increasing
$\theta$ alters the shape of the higher-order image and obscures the
horizon's outline. Moreover, as the observation frequency increases,
the corresponding observed intensity decreases, making the
higher-order image, and the horizon boundary is clearly visible.
However, the position of the higher-order image remains constant for
all observed frequencies, because the gravitational lensing effect
of the black hole does not depend on the frequency of light.

In the scenario of anisotropic radiation under the RIAF model, the
impact of $\lambda$ and $\theta$ on the black hole shadow remains
the same as that under isotropic radiation. However, differences
from isotropic radiation are that the higher-order images
significantly deform into an ellipsoidal shape, which remains
approximately circular in the case of isotropic radiation at $\theta
= 80^\circ$. The intensity profiles of the direct image under
anisotropic radiation are significantly greater than under isotropic
radiation. Additionally, the intensity of the ring structure in the
vertical direction is significantly greater than in the horizontal
direction. This is in stark contrast to isotropic radiation
scenario, where the horizontal intensity is more pronounced than the
vertical intensity. All these differences are more visible at higher
inclinations. Comparing the results of observed frequencies, it is
noticed that, with the increasing values of the observed frequency,
the intensity distribution shrinks to a narrow region, similar to
the case of isotropic radiation. However, unlike isotropic
radiation, under anisotropic radiation, the maximum intensity in the
vertical direction is consistently larger than that in the
horizontal direction.

Next, we investigated the imaging and polarization properties of the
BAAF disk model. For the distribution of intensity profiles, we
again considered purely infalling motion of accreting matter with
anisotropic synchrotron emission, for a direct comparison with the
RIAF model. In this case, with the increase of both $\lambda$ and
$\theta$, results in a reduction of the size of the higher-order
images, and enhances the brightness of the direct images beyond the
higher-order images, respectively, but hardly changes the size of
the higher-order images, which are in sharp contrast to the
phenomenological model. To intuitively compare the differences
between the aforementioned two models, it is noticed that at the
same observation position, the direct image in the phenomenological
model is significantly brighter as compared to the Hou disk model.
At higher inclinations, the obscuration effect outside the
equatorial plane radiation on the event horizon contour is more
visible in the phenomenological model, indicating that the
gravitational lensing effect is stronger in the phenomenological
model. And the observational frequency profiles exhibit the similar
results as discussed in the RIAF model.

We subsequently examined the polarization characteristics of the
images produced by the BAAF model. The polarized intensity closely
align with the total intensity distribution, showing stronger
polarization in regions of higher brightness. Both the polarization
degree and the direction display a pronounced dependence on the GB
parameter $\lambda$ as well as on the observer's position, reflects
the intrinsic structure of spacetime. Importantly, unlike in thin
disk model where the inner shadow remains unpolarized, while in
geometrically thick disk models, the lensing of off-equatorial
emission allows polarization vectors to spread throughout the whole
image plane.

To conclude this work, we noticed that the model parameters play a
major role in shaping shadow images in the framework of
geometrically thick disk models. Compared with geometrically thin
disks, geometrically thick and optically thin accretion flows
provide a more realistic display of astrophysical environments. By
combining intensity and polarization features, thick disk models
offer a more comprehensive probe of the radiation characteristics
and the underlying spacetime geometry surrounding black holes. As
future work, it is expected to extend the above analysis for the
Kerr-like black hole in the EGB gravity framework.\\
{\bf Acknowledgements}\\
This work is supported by the National Natural Science Foundation of
China (Grants Nos. 12375043, 12575069 ).

\end{document}